\documentclass[11pt,a4paper]{article}
\pdfoutput=1
\usepackage{jheppub}
\usepackage{cite}
\usepackage{physics}
\usepackage[normalem]{ulem}
\usepackage{braket}
\usepackage{subcaption}

\usepackage{multirow, graphicx,amssymb,url,mathrsfs,amsmath}
\usepackage{wrapfig,boxedminipage,epsfig}
\usepackage{amsxtra,amstext,latexsym,dsfont,amsfonts}
\usepackage{color}
\usepackage[dvipsnames]{xcolor}
\usepackage{float}
\usepackage{slashed}
\usepackage{calligra}
\usepackage{enumerate,ulem}
\DeclareFontShape{T1}{calligra}{m}{n}{<->s*[2.2]callig15}{}
\DeclareMathAlphabet{\mathcalligra}{T1}{calligra}{m}{n}

\newcommand{\be}{\begin{equation}}
\newcommand{\ee}{\end{equation}}
\newcommand{\bea}{\begin{eqnarray}}
\newcommand{\eea}{\end{eqnarray}}

\newcommand\GM[1][]{{\rm GM}^{\IfInteger{#1}{(#1)}{\mathtt{(#1)}}}}

\title{Genuine multi-entropy and holography}

 \author[a, b]{Norihiro Iizuka}
 
 \author[c]{and Mitsuhiro Nishida}
 
\affiliation[a]{\it Department of Physics, National Tsing Hua University, Hsinchu 300044, Taiwan}

\affiliation[b]{\it Yukawa Institute for Theoretical Physics, Kyoto University, Kyoto 606-8502, Japan}

\affiliation[c]{\it National Institute of Technology, Yuge College, Ehime 794-2593, Japan}
 
\emailAdd{iizuka@phys.nthu.edu.tw}
\emailAdd{mnishida124@gmail.com}

\abstract{
Is bipartite entanglement sufficient for holography? Through the analysis of the Markov gap \cite{Akers:2019gcv, Hayden:2021gno}, it is known that the answer is no. In this paper, we give a new perspective on this issue from a different angle using a multi-entropy \cite{Gadde:2022cqi,Penington:2022dhr}. 
We define a genuine $\mathtt{q}$-partite multi-entropy from a $\mathtt{q}$-partite multi-entropy by subtracting appropriate linear combinations of $\mathtt{\tilde{q}}$-partite multi-entropies for $\mathtt{\tilde{q}} < \mathtt{q}$, in such a way that the genuine $\mathtt{q}$-partite multi-entropy vanishes for all $\mathtt{\tilde{q}}$-partite entangled states. After studying several aspects, we apply it to black holes and holography. For the application to black holes, we see that such a genuine $\mathtt{q}$-partite multi-entropy is important only after the Page time.  For the application to holography, we prove that non-bipartite multi-entropies are always positive and $\mathcal{O}\left({1/ G_N}\right)$, as long as boundary subregions are connected. This indicates that for holography, genuine multi-partite entanglement is not small and plays an important role. 
}

\begin{document}
\maketitle

\section{Introduction}

Since the Ryu-Takayanagi (RT) formula \cite{Ryu:2006bv}, which relates quantum entanglement entropy between boundary regions into minimal geometrical surfaces (RT-surface) in the bulk, was proposed,  
significant progress has been made in our understanding of holography from a quantum information viewpoint.  Random tensor network (RTN) \cite{Hayden:2016cfa} is one of such interesting development, which shows a very analogous formula to RT for the entanglement entropy and this analogy gives the view that RTN is a toy model of quantum gravity. 
Another interesting picture is given by a bit thread picture \cite{Freedman:2016zud}, which gives a new perspective to the entanglement entropy as the maximum number of `bit threads'. 
Given that entanglement entropy is a bipartite measure, it is natural to ask if bipartite entanglement is enough for holography and it is conjectured that the entanglement of holographic CFT states is mostly bipartite (mostly-bipartite conjecture) in \cite{Cui:2018dyq}. 

In general quantum systems, one can divide the total system into three or more subsystems, and many states are involved in entanglement. In general, there is more than bipartite entanglement involved.
For example, in spin systems, generic states contain more than bipartite entanglement, {\it i.e.,} they contain genuine multi-partite entanglement such as tripartite and quadripartite entanglement.
Famous examples of tripartite entanglement are the GHZ state and the W-state \cite{Dur:2000zz}, which are distinguished from bipartite entangled states. 

Even though the mostly-bipartite conjecture is quite interesting, it was disproved through the study of the Markov gap\footnote{We review the argument of \cite{Hayden:2021gno} in footnote \ref{posimarkovproof} at the end of section \ref{sec:proof}, and compare it with our holographic proof.}  \cite{Akers:2019gcv, Hayden:2021gno}. Markov gap is defined as the difference between reflected entropy \cite{Dutta:2019gen} and mutual information.
The Markov gap is a useful quantity to detect genuine tripartite entanglement contribution in holography since it vanishes for all states that contain only bipartite entanglement. Recently, another new multi-partite quantum entanglement measure has been proposed, which is called the multi-entropy \cite{Gadde:2022cqi,Penington:2022dhr,Gadde:2023zzj}. In this paper, we would like to obtain a new perspective on the above-mentioned point by defining a {\it genuine} $\mathtt{q}$-partite multi-entropy using the multi-entropy. This is because the multi-entropy by itself can be nonzero even by a bipartite contribution only. On the other hand, the genuine $\mathtt{q}$-partite multi-entropy we define can become nonzero {\it only if} the state contains genuine $\mathtt{q}$-partite entanglement. By genuine $\mathtt{q}$-partite entanglement, we mean the entanglement that cannot be decomposed into tensor products of lower-partite ({\it i.e.,} less than $\mathtt{q}$-partite) entanglement. We emphasize that the genuine multi-entropy can be defined for any integer $\mathtt{q}\ge3$. Note that how to construct such a quantity is quite nontrivial, and in this paper, we give a prescription for how to construct the genuine multi-entropy with a new $\mathtt{q}=4$ concrete example. We also point out the analogy between the Markov gap and genuine multi-entropy for $\mathtt{q}=3$. Markov gap is defined using reflected entropy. On the other hand, genuine multi-entropy is defined using multi-entropy. We will see more of their analogy in detail\footnote{Other than the Markov gap, in \cite{Ju:2024hba,Ju:2024kuc} aspects of holographic multipartite entanglement are studied. In \cite{Mori:2024gwe}, the absence of distillable bipartite entanglement in some holographic regions is suggested. See \cite{Li:2025nxv} as well.}. 

The outline of this paper is as follows. 
In section \ref{sec:GMEq3}, after reviewing the definition of a multi-entropy and R\'enyi multi-entropy, we define the genuine $\mathtt{q}$-partite multi-entropy for $\mathtt{q}=3$ case as a linear combination of the $\mathtt{q}=3$ and $\mathtt{q}=2$ multi-entropies. We also study the genuine multi-entropy for $\mathtt{q}=3$ of two qubit states, namely the GHZ and W states.
In section \ref{sec:BHGMEC}, we use this genuine $\mathtt{q}=3$ multi-entropy to study genuine tripartite entanglement in an evaporating black hole.
The crucial approximation here is that we approximate an evaporating black hole and its radiation with a Haar-random state. 
Then we divide the total system into three subsystems, two for Hawking radiation subsystems and one for an evaporating black hole subsystem.
By computing the genuine $\mathtt{q}=3$ multi-entropy 
in random states, we observe the time scale when the effect of genuine tripartite entanglement becomes large is always after the Page time. 
In section \ref{sec:proof}, we also apply the genuine multi-entropy in holography. 
We give a simple proof that non-bipartite, generic $\mathtt{q}$ entanglement is always positive and of order $\mathcal{O}(N^2)=\mathcal{O}(1/G_N)$ in holographic settings as far as the boundary subregions are connected. This explains why bipartite entanglement is not enough in holography, and this is similar to the nonnegative Markov gap \cite{Akers:2019gcv, Hayden:2021gno}. 
In this paper, we start a genuine $\mathtt{q}$-partite multi-entropy from $\mathtt{q}=3$ tripartite genuine case and also define genuine multi-entropy for $\mathtt{q}=4$. However, we would like to stress that one can define genuine multi-entropy for any $\mathtt{q}$ as well. In section \ref{sec:GMEq4}, we discuss the genuine multi-entropy for $\mathtt{q} \ge 4$. We also show the black hole genuine multi-entropy curves for $\mathtt{q}=4$ and see that just as $\mathtt{q}=3$, the time scale when the effect of genuine quadripartite entanglement becomes large is always after the Page time. After briefly giving the prescription on how to construct on genuine multi-entropy, we end in section \ref{sec:discussion} with open questions. 
Appendix \ref{App:MaxMEW} is for the maximum value of the genuine tri-partite multi-entropy in W-class, and Appendix \ref{App:TripartiteNegativity} is for other measures constructed as products of bipartite measures. 
Appendix \ref{App:CforIq4} is for the evaluation of the $\mathtt{q}$-partite information $I_{\mathtt{q}}$ curves for black hole evaporation. 

{\bf Note added:} After writing the draft, we learned that $\mathtt{q}=3$ part of the results in section \ref{sec:proof} overlap with \cite{Harper:2024ker}.

\section{Defining a genuine multi-entropy in tripartite systems}\label{sec:GMEq3}
In this section, we introduce the definition of multi-entropy and  {\it genuine} tripartite multi-entropy by the linear combination of multi-entropies.  Before we proceed, we first give our criterion for `genuineness'. 

\begin{figure}[t]
    \centering
    \includegraphics[width=6cm]{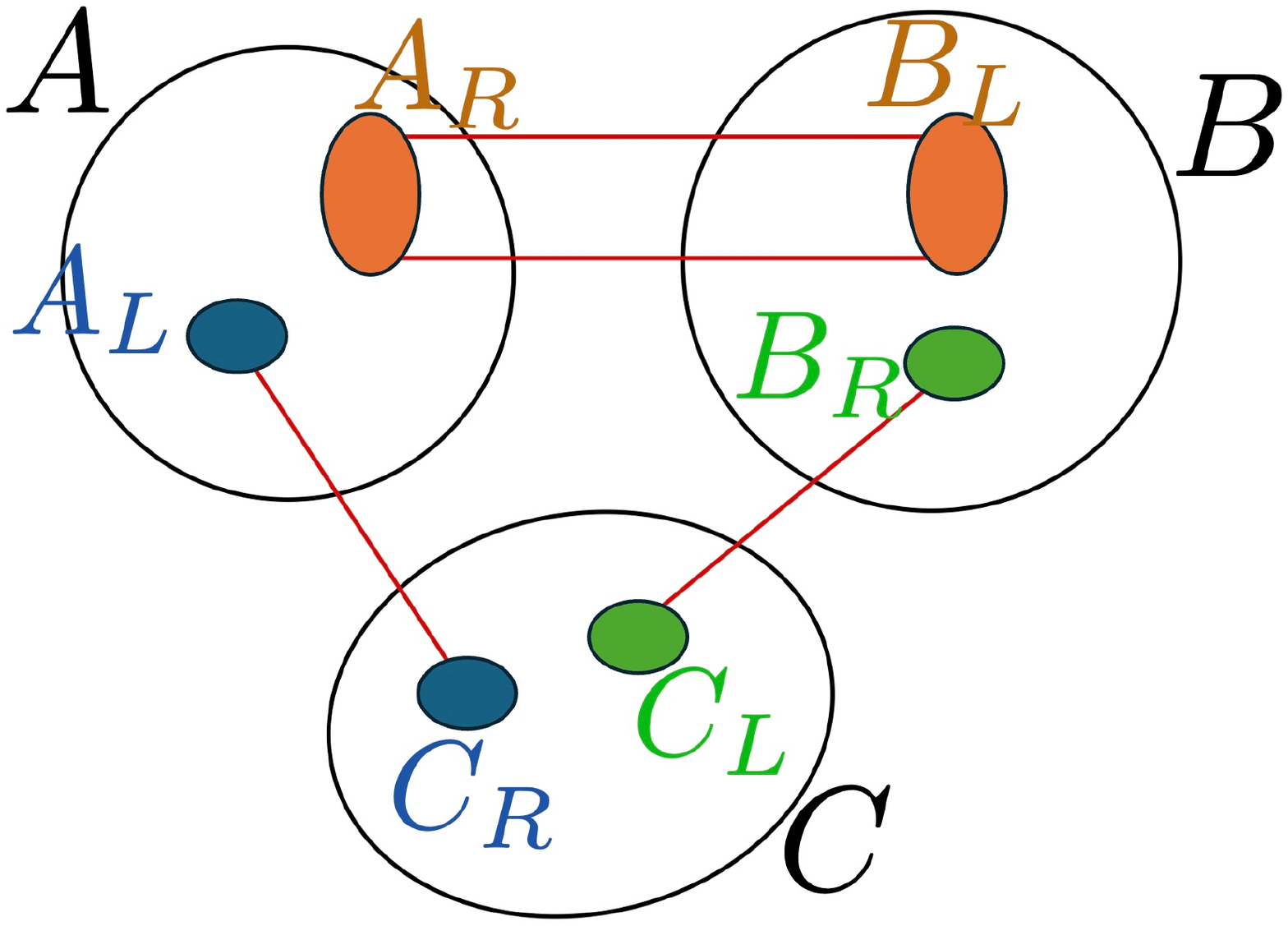}
    \caption{Triangle state is composed of bipartite entangled states only. Red lines represent bipartite entanglements.}
    \label{fig:triangle}
\end{figure}

For that purpose, let us first consider a state that contains {\it only} bipartite entanglement. A typical example of such states is a triangle state $\ket{\Delta}_{ABC}$, which is defined in \cite{Zou:2020bly} as 
\begin{align}
\ket{\Delta}_{ABC}= \ket{\psi}_{A_RB_L}\ket{\psi}_{B_RC_L}\ket{\psi}_{C_RA_L} \,,
\label{eq:triangle}
\end{align}
for given three subsystems $A$, $B$, $C$.
See Fig.~\ref{fig:triangle}. 
{\it Genuine} tripartite entanglement measures must be zero for all of the triangle states. 
Later in section \ref{defsection}, we define the genuine tripartite multi-entropy, which we denote ${\rm GM}^{(3)}(A:B:C)$, to satisfy this genuineness. In other words, our genuine tripartite multi-entropy, ${\rm GM}^{(3)}(A:B:C)$, excludes all bipartite entanglement contributions. Similarly we define a genuine $\mathtt{q}$-partite multi-entropy, which we denote ${\rm GM}^{(q)}(A_1:A_2:\cdots : A_\mathtt{q})$, as a quantity that vanishes for all states which can be written as tensor products of entangled subsystems involving fewer than $\mathtt{q}$ parties.

The Markov gap, defined as $S_R - I$ in \cite{Akers:2019gcv, Hayden:2021gno}, also satisfies this criterion \cite{Zou:2020bly}; that is, it vanishes for all states that exhibit only bipartite entanglement, and therefore serves as an indicator for the presence of tripartite entanglement. Here $S_R$ is the reflected entropy\footnote{Suppose a pure state $\rho_{ABC}$ is defined on a tripartite system consisting of subsystems $A$, $B$, $C$. The reflected entropy $S_R$ for the pair $A$ and $B$ is defined as follows \cite{Dutta:2019gen}: First, trace out subsystem $C$ to obtain a reduced density matrix $\rho_{AB} = \Tr_C \rho_{ABC}$. Then canonically purify it to $\rho_{AA^*BB^*}$, where $A^*$ and $B^*$ are auxiliary purifying systems for $A$ and $B$, respectively. The reflected entropy is given by the entanglement entropy between $AA^*$ and $BB^*$, namely the von Neumann entropy of the $\rho_{AA^*}  = \Tr_{B B^*} \rho_{AA^*BB^*}$.} and $I$ is the mutual information\footnote{The mutual information is simply $I(A,B) = S(A) + S(B) - S(AB) = S(A) + S(B) - S(C)$, if the total system $ABC$ is pure.}. Although both our genuine multi-entropy ${\rm GM}^{(3)}$ and the Markov gap vanish for states with only bipartite entanglement, they differ in their construction: the Markov gap is defined via the reflected entropy, whereas the genuine tripartite multi-entropy is defined in terms of the multi-entropy. We will comment further on their similarities and differences later, but before doing so, we briefly review the definitions of multi-entropy and Rényi multi-entropy.

\subsection{Multi-entropy in tripartite systems}
\begin{figure}[t]
    \centering
    \includegraphics[width=10cm]{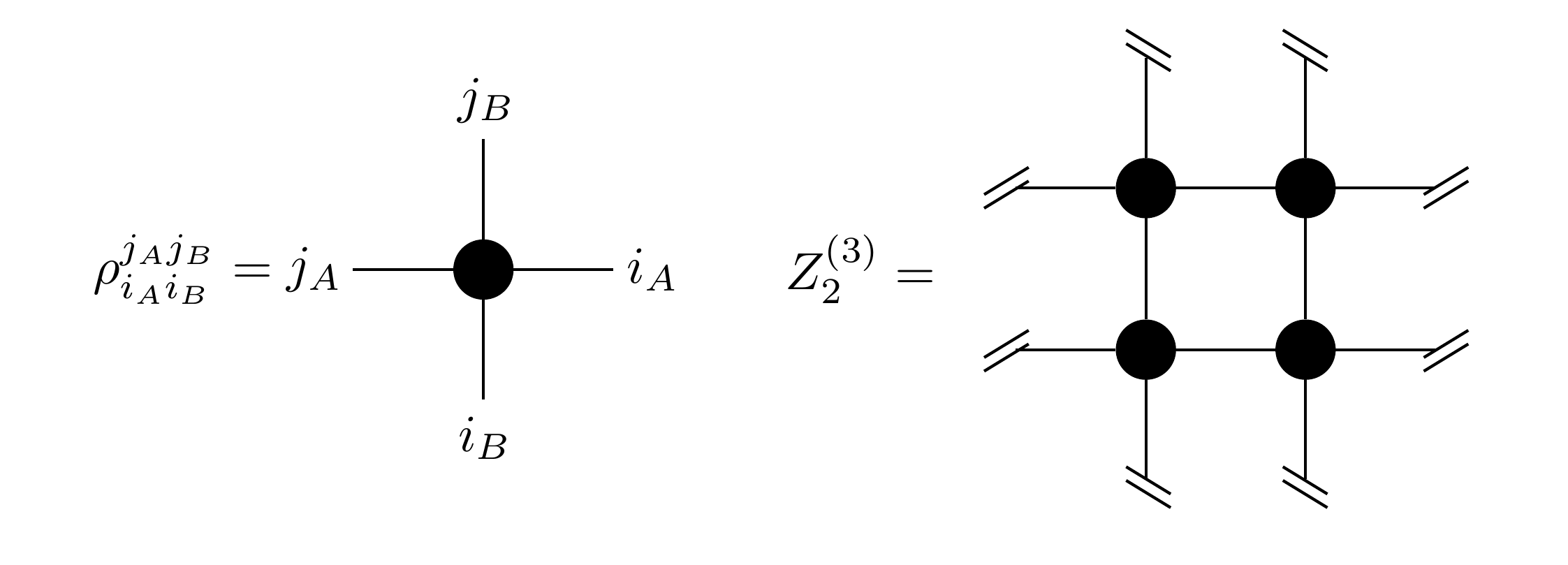}
    \caption{Contraction pattern of a reduced density matrix $\rho=\text{Tr}_C\ket{\psi}\bra{\psi}$ for $Z_2^{(3)}$.}
    \label{fig:Z23}
\end{figure}

The R\'enyi mutli-entropy \cite{Gadde:2022cqi,Penington:2022dhr, Gadde:2023zzj} was defined as a symmetric multi-partite entanglement measure in $\mathtt{q}$-partite systems. For instance, let us consider a tripartite pure state $|\psi\rangle$ on $\mathcal{H}=\mathcal{H}_A\otimes\mathcal{H}_B\otimes\mathcal{H}_C$. For such $\mathtt{q}=3$ tripartite systems, R\'enyi multi-entropy \(S^{(3)}_n(A:B:C)\) of $|\psi\rangle$ is defined by\footnote{This definition of multi-entropy follows \cite{Penington:2022dhr}, which is different from the definition \cite{Gadde:2022cqi} by a factor $1/n$. Thus, for generic $\mathtt{q}$, there are factor $1/n^{\mathtt{q}-2}$ difference \cite{Gadde:2023zni}. In our previous paper \cite{Iizuka:2024pzm}, we define the multi-entropy without this $1/n^{\mathtt{q}-2}$ factor \label{fn1}.}
\begin{align}
S^{(3)}_n(A:B:C) &:= \frac{1}{1-n}\frac{1}{n}\log \frac{Z^{(3)}_n}{(Z^{(3)}_1)^{n^{2}}},\label{DefRME}\\
Z^{(3)}_n &:= \bra{\psi}^{\otimes n^{2}} \Sigma_A(g_A)\Sigma_B(g_B)\Sigma_C(g_C)\ket{\psi}^{\otimes n^{2}},
\end{align}
where \(\Sigma_{A,B,C}\) are the twist operators for the permutation action of \(g_{A,B,C}\) on indices of density matrices. Their explicit representation of the permutation group is
\begin{align}
g_A&=(1,2,\dots,n)(n+1,n+2,\dots,2n)\cdots(n^2-n+1,n^2-n+2,\dots, n^2),\\
g_B&=(1,n+1,\dots,n^2-n+1)(2,n+2,\dots,n^2-n+2)\cdots(n,2n,\dots, n^2),\\
g_C&=(1)(2)\cdots(n^2).
\end{align}
See Fig.~\ref{fig:Z23}. 
The multi-entropy $S^{(3)}(A:B:C)$ is defined by 
\begin{align}
S^{(3)}(A:B:C):=\lim_{n\to1}S^{(3)}_n(A:B:C).
\end{align}

One can understand this multi-entropy as a generalization of the purity $\Tr(\rho^2)$, as follows.
Consider the case where subsystem $B$ is one-dimensional. Then, in Fig.~\ref{fig:Z23}, the indices $i_B$ and $j_B$ (represented by vertical lines) become trivial since $i_B = j_B = 1$. In this limit, $Z_2^{(3)}$ reduces to two copies of the purity of the reduced density matrix for subsystem $A$, {\it i.e.}, $Z_2^{(3)} \to \left( \Tr(\rho_A^2) \right)^2$.
Similarly, if instead subsystem $A$ is one-dimensional, the horizontal indices $i_A$ and $j_A$ lose their meaning, and $Z_2^{(3)}$ reduces to two copies of the purity of $\rho_B$: $Z_2^{(3)} \to \left( \Tr(\rho_B^2) \right)^2$.

Note that the purity-type contractions for subsystems $A$ and $B$ are taken along the horizontal and vertical directions, respectively.
More generally, for a tripartite density matrix $\rho_{ABC}$, after tracing out subsystem $C$, we compute purity-type contractions for $A$ along horizontal lines and for $B$ along vertical lines. This procedure yields the $\mathtt{q} = 3$ and $n=2$ R\'enyi multi-entropy represented in Fig.~\ref{fig:Z23}. 

This construction can be generalized to arbitrary $\mathtt{q}$ and $n$: by performing $n$-fold purity-type contractions $\Tr(\rho^n)$ along $\mathtt{q} - 1$ mutually orthogonal directions, thus we have $n^{\mathtt{q}-1}$ replicas and in this way, one obtains the R\'enyi multi-entropy $S_n^{(\mathtt{q})}$.

The R\'enyi multi-entropy has the following properties \cite{Penington:2022dhr, Gadde:2023zni}.
\begin{itemize}
\item $S^{(3)}_n(A:B:C)$ is invariant under local unitary transformations of $|\psi\rangle$. 
\item $S^{(3)}_n(A:B:C)$ is symmetric in all the parties of tripartite systems:
\begin{align}
&\;\;\;\;\;S^{(3)}_n(A:B:C)=S^{(3)}_n(A:C:B)=S^{(3)}_n(B:A:C)\notag\\
&=S^{(3)}_n(B:C:A)=S^{(3)}_n(C:A:B)=S^{(3)}_n(C:B:A).\label{symmetric}
\end{align}
\item $S^{(3)}_n(A:B:C)$ is additive under tensor products of pure states. If $|\psi\rangle_{A_1B_1C_1}$ and $|\phi\rangle_{A_2B_2C_2}$ are pure states, $S^{(3)}_n(A_1A_2:B_1B_2:C_1C_2)_{|\psi\rangle\otimes|\phi\rangle}$ of $|\psi\rangle_{A_1B_1C_1}\otimes|\phi\rangle_{A_2B_2C_2}$ is given by
\begin{align}
\hspace{-4mm}S^{(3)}_n(A_1A_2:B_1B_2:C_1C_2)_{|\psi\rangle\otimes|\phi\rangle}=S^{(3)}_n(A_1:B_1:C_1)_{|\psi\rangle}+S^{(3)}_n(A_2:B_2:C_2)_{|\phi\rangle}.\label{additive}
\end{align}
\end{itemize}
These properties also hold for $(\mathtt{q},n)$-R\'enyi multi-entropy in $\mathtt{q}$-partite systems.
As we will see soon, the additive property given by \eqref{additive} plays an important role in defining genuine multi-entropy.

As concrete examples, we consider $S^{(3)}_n$ of W-state and GHZ state in three-qubit systems, where W-state and GHZ state \cite{Dur:2000zz} are given by
\begin{align}
    |\text{GHZ}\rangle&:=\frac{1}{\sqrt{2}}\left(|000\rangle+|111\rangle\right),\\
    |\text{W}\rangle&:=\frac{1}{\sqrt{3}}\left(|100\rangle+|010\rangle+|001\rangle\right).
\end{align}
For small $n=2,3,4$, $S^{(3)}_n$ of W-state is evaluated in 
\cite{Gadde:2022cqi} but generic $n$ formula for R\'enyi multi-entropy is not known. They behave: 
\begin{align}
\label{Wseries}
&S^{(3)}_{2 \; {|\text{W}\rangle}}=\frac{1}{2}\log 9 = \log 3,\\ \;
&S^{(3)}_{3 \; {|\text{W}\rangle}}=\frac{1}{6}\log \left(\frac{6561}{14}\right) \approx \frac{1}{6} \log \frac{3^8}{3^{2.4}} \approx 0.93 \log 3,\\
&S^{(3)}_{4 \; {|\text{W}\rangle}}=\frac{1}{12}\log \left(\frac{14348907}{139}\right) \approx \frac{1}{12} \log \frac{3^{15}}{3^{4.5}} \approx {0.88} \log 3,\;\\
&S^{(3)}_{5 \; {|\text{W}\rangle}}=\frac{1}{20}\log \left(\frac{94143178827}{847}\right) \approx \frac{1}{20} \log \frac{3^{23}}{3^{6.1}}  \approx 0.84 \log 3,\; \\
&{S^{(3)}_{6 \; {|\text{W}\rangle}}=\frac{1}{30}\log \left(\frac{50031545098999707}{87880}\right) \approx \frac{1}{30} \log \frac{3^{35}}{3^{10.4}}  \approx 0.82 \log 3.}\;
\label{RMEW}
\end{align}
From these, we conjecture that 
\begin{align}
\label{Wconjecture}
S^{3}_{1 \; {|\text{W}\rangle}}= c_{\text{W}} \log 3 \,, \quad \mbox{where $c_{\text{W}}  > 1$}  \,.
\end{align}

On the other hand, for GHZ state, generic $n$ dependence for R\'enyi multi-entropy is known 
\begin{align}
S^{(3)}_{2 \; {|\text{GHZ}\rangle}}&=\frac{3}{2}\log2, \;\quad \,\,\,\, S^{(3)}_{3 \; {|\text{GHZ}\rangle}}=\frac{4}{3}\log2,  \;\quad \,\,\,\, S^{(3)}_{4 \; {|\text{GHZ}\rangle}}=\frac{5}{4}\log2,\label{RMEGHZ}\\
S^{(3)}_{n \; {|\text{GHZ}\rangle}}&=\frac{1+n}{n}\log2, \; \,\, \,  S^{(3)}_{{|\text{GHZ}\rangle}}=2\log2.
\end{align}

As another example, consider a triangle state $\ket{\Delta}$ given by \eqref{eq:triangle}, 
which has no genuine tripartite entanglement. For such a triangle state, its R\'enyi multi-entropy $S^{(3)}_n(A:B:C)_{|\Delta\rangle}$ satisfies the following relationship \cite{Penington:2022dhr}
\begin{align}
S^{(3)}_n(A:B:C)_{|\Delta\rangle}=\frac{1}{2}\left(S_n^{(2)}(A)_{|\Delta\rangle}+S_n^{(2)}(B)_{|\Delta\rangle}+S_n^{(2)}(C)_{|\Delta\rangle}\right),\label{RMETState}
\end{align}
where 
\be
S_n^{(2)}(A)_{|\Delta\rangle}:=\frac{1}{1-n}\log \text{Tr}_A\rho_A^n
\ee is a R\'enyi entanglement entropy for reduced density matrix $\rho_A:=\text{Tr}_{BC}|\Delta\rangle\langle\Delta|$. 
Thus, we also write $S_n^{(2)}(A:BC) = S_n^{(2)}(A)$, $S_n^{(2)}(B:CA) = S_n^{(2)}(B)$, $S_n^{(2)}(C:AB)= S_n^{(2)}(C)$. 
In particular, for three-qubit systems $A$, $B$, $C$, we obtain 
\begin{align}
S^{(3)}_n(A:B:C)_{|\Delta\rangle}\le\log 2,
\end{align}
which is smaller than $S^{(3)}_{n \; {|\text{GHZ}\rangle}}$ if $n>0$.

By comparing (\ref{RMEW}) and (\ref{RMEGHZ}), one can see that $S^{(3)}_{n \; {|\text{W}\rangle}}$ is larger than $S^{(3)}_{n \; {|\text{GHZ}\rangle}}$ in the calculation examples above for $n\ge 2$. In addition, our numerical computations imply that $S^{(3)}_{2 \; {|\text{W}\rangle}}$ is maximum among $S^{(3)}_{2}$ in three qubit systems. As a partial proof, in Appendix \ref{App:MaxMEW}, we show that $S^{(3)}_{2 \; {|\text{W}\rangle}}$ is maximum among $S^{(3)}_{2}$ in W-class. Therefore, we conjecture that $S^{(3)}_{n \; {|\text{W}\rangle}}$ is maximum among $S^{(3)}_{n}$ in three qubit systems.

\subsection{Genuine multi-entropy in tripartite systems}
\label{defsection}
Generally, $S^{(3)}_n$ is nonzero for pure states with bipartite entanglement as seen in \eqref{RMETState}. In tripartite systems, there are two types of entanglement: bipartite entanglement between two subsystems and genuine tripartite entanglement between three subsystems \cite{Dur:2000zz}. 

For generic $\mathtt{q}$, we define a genuine $\mathtt{q}$-partite multi-entropy from a $\mathtt{q}$-partite multi-entropy by subtracting appropriate linear combinations of $\mathtt{\tilde{q}}$-partite multi-entropies for $\mathtt{\tilde{q}} < \mathtt{q}$, in such a way that the genuine $\mathtt{q}$-partite multi-entropy vanish for all $\mathtt{\tilde{q}}$-partite entangled states. Thus, for $\mathtt{q}=3$ case, we define the genuine multi-entropy in such a way that it gives zero for pure states with only bipartite entanglement. Then the following linear combination works, 
\begin{align}
\label{def_n_GM}
{\rm GM}_n^{(3)}(A:B:C) 
&:=S^{(3)}_n(A:B:C)-\frac{1}{2}\left(S_n^{(2)}(A:BC)+S_n^{(2)}(B:CA)+S_n^{(2)}(C:AB)\right).
\end{align}
From (\ref{RMETState}), it immediately follows that ${\rm GM}_n^{(3)}(A:B:C)$ of the triangle state $|\Delta\rangle$, which has only bipartite entanglement, is zero.
We call ${\rm GM}_n^{(3)}(A:B:C)$ genuine $n$-th R\'enyi multi-entropy $(\mathtt{q}=3)$ in tripartite systems. 
As an independent work from us, in \cite{Liu:2024ulq} the same quantity \eqref{def_n_GM} is defined as $\kappa_n$ and values on 2d CFTs are studied.

We define the genuine $\mathtt{q}=3$ multi-entropy ${\rm GM}^{(3)}(A:B:C)$ by taking the $n\to 1$ limit as 
\begin{align}
\label{n1limit}
{\rm GM}^{(3)}(A:B:C):=\lim_{n\to1}{\rm GM}_n^{(3)}(A:B:C).
\end{align}
Later in section \ref{sec:proof}, we prove that 
\begin{align}
\label{inequality}
{\rm GM}^{(3)}(A:B:C) \ge 0 
\end{align}
using holographic duals. 

The R\'enyi  $n=2$ case is interesting as well since in this case, the multi-entropy $S^{(3)}_{n=2}(A:B:C)$ essentially reduces to the reflected entropy. See Figure 13 of \cite{Iizuka:2024pzm} for the four replica contraction diagram. In this case, one can show that the genuine $n=2$ R\'enyi multi-entropy $(\mathtt{q}=3)$ is related to the R\'enyi Markov gap, which is the difference between the R\'enyi reflected entropy and the R\'enyi mutual entropy, as \cite{Liu:2024ulq}
\be
{\rm GM}_{n=2}^{(3)}(A:B:C) \propto \mbox{(($m=2,n=2$) R\'enyi Markov gap)}.
\ee
We will comment on this relation more in detail in the next section.

Note that both the Markov gap and the genuine multi-entropy $\mathtt{q}=3$ vanish for the states containing only bipartite entanglement. Thus for $\mathtt{q}=3$ tripartite case, their properties are expected to be similar in general. Next, we will comment on properties of the genuine $\mathtt{q} = 3$ multi-entropy through explicit examples. 

For W-state, one can obtain ${\rm GM}_{n}^{(3)}$ for small $n$ as follows, 
\begin{align}
{\rm GM}^{(3)}_{2 \; {|\text{W}\rangle}}&=\log \left(\frac{5 \sqrt{5}}{9}\right),\; \qquad \quad \,\,\,  {\rm GM}^{(3)}_{3 \; {|\text{W}\rangle}}=\frac{1}{12} \log \left(\frac{2187}{196}\right), \; \\
{\rm GM}^{(3)}_{4 \; {|\text{W}\rangle}}&=\frac{1}{12} \log \left(\frac{24137569}{2735937}\right), \; \, {\rm GM}^{(3)}_{5 \; {|\text{W}\rangle}}=\frac{1}{40} \log \left(\frac{285311670611}{234365481}\right),\\
{{\rm GM}^{(3)}_{6 \; {|\text{W}\rangle}}}&=\frac{1}{120} \log\left(\frac{72890483685103052142902866787761839379440139451}{72512083137473374754661832057671
   9360000}\right),
\label{GRMEW}
\end{align}
using 
\begin{align}
\label{Wbipartite}
S^{(2)}_{n \; {|\text{W}\rangle}}&=\frac{\log \left(\left(2^n+1\right)/3^n\right)}{1-n} \,, \quad   \mbox{with} \quad S^{(2)}_{ \; {|\text{W}\rangle}}=  \log 3 - \frac{2}{3} \log 2 \,.
\end{align}
All of  ${\rm GM}_{n}^{(3)}$ for $n=2, 3, 4, 5, 6$ are positive. Finally 
using \eqref{Wconjecture} and \eqref{Wbipartite}, for $n\to1$ case we have  
\begin{align}
{\rm GM}^{(3)}_{ \; {|\text{W}\rangle}}&=\log \left( 2 \, 3^{c_W - \frac{3}{2}} \right) > \log \left( 2 \, 3^{ - \frac{1}{2}} \right) \approx 0.14 >  0 \quad \mbox{for $c_W > 1$.}
\label{GM3W}
\end{align}
This is the new result as far as we are aware.

 In contrast, ${\rm GM}^{(3)}_n$ of GHZ state is \cite{Liu:2024ulq}
\begin{align}
{\rm GM}^{(3)}_{n \; {|\text{GHZ}\rangle}} &=\left(\frac{1+n}{n}-\frac{3}{2}\right)\log2.\,
\end{align}
Especially this means for $n\to 1 $ and $n\to 2$ cases, 
\begin{align} 
\label{GM3GHZ}
{\rm GM}^{(3)}_{ {|\text{GHZ}\rangle}} &=\frac{1}{2}\log2  > 0 \,\quad (n \to 1) \\ 
{\rm GM}^{(3)}_{2 \; {|\text{GHZ}\rangle}}&=0 \quad (n \to 2),
\end{align}
where genuine multi-entropy ${\rm GM}^{(3)}_{ {|\text{GHZ}\rangle}}$ is positive, but ${\rm GM}^{(3)}_{n \; {|\text{GHZ}\rangle}}$ is negative if $n>2$.
Thus, even though ${\rm GM}^{(3)}$ is expected to behave similarly to Markov gap \cite{Hayden:2021gno}, there are important differences. One difference between Markov gap and genuine multi-entropy ${\rm GM}^{(3)}$ is that Markov gap of GHZ state is zero, but ${\rm GM}^{(3)}$ of GHZ state is nonzero as \eqref{GM3GHZ}. This means that the Markov gap cannot capture genuine tripartite entanglement in the GHZ state, on the other hand, ${\rm GM}^{(3)}$ can capture it. Another note is that even though we have not yet proved it, it is likely that $c_W > 1$ in \eqref{Wconjecture} from \eqref{Wseries} - \eqref{RMEW}. This implies that for both GHZ state and W state,  ${\rm GM}^{(3)} > 0$ as \eqref{GM3W} and \eqref{GM3GHZ}.  This inequality is the same as \eqref{inequality}, however the origin is totally different since there is no guarantee that GHZ state or W state can be holographic.

So far, we focus on $\mathtt{q}=3$ tripartite case. However similarly it is possible to define genuine multi-entropy for higher $\mathtt{q}$ cases as well, using the same principles. We will show this more explicitly for $\mathtt{q}=4$ case in Section \ref{sec:GMEq4}.

Other than the simple spin systems, several papers studied 2d CFT cases as in \cite{Penington:2022dhr,Gadde:2023zzj, Harper:2024ker}. In particular, ${\rm GM}^{(3)}_n(A:B:C)$ in 2d CFTs and a free fermion model on a 2d square lattice was studied in  \cite{Liu:2024ulq}. See also \cite{Yuan:2024yfg, Gadde:2024taa} for 2d CFT computations for some generalizations of the multi-entropy.

Before we proceed, we comment on other proposal for the genuine multi-partite entanglement. One of such proposal is the `L-entropy' defined by  \cite{Basak:2024uwc}. 
The bipartite L-entropy $l_{AB}$ was defined as 
\be
l_{AB} := \text{Min}\{2S^{(2)}(A),2S^{(2)}(B)\} - S^{(1,1)}_R(A:B),
\ee
which is motivated by the inequality $S^{(1,1)}_R(A:B)\le\text{Min}\{2S^{(2)}(A),2S^{(2)}(B)\}$, just as the Markov gap was motivated by the inequality $I^{(1)}(A:B)\le S^{(1,1)}_R(A:B)$. The tripartite L-entropy is a product of three bipartite L-entropies, namely $l_{AB}, l_{BC}, l_{CA}$. However, the L-entropy does not satisfy the additive property \eqref{additive} due to  
\begin{align}
    &\text{Min}\{S^{(2)}(A_1),S^{(2)}(B_1)\}+\text{Min}\{S^{(2)}(A_2),S^{(2)}(B_2)\} \nonumber \\
    &\ne\text{Min}\{S^{(2)}(A_1)+S^{(2)}(A_2),S^{(2)}(B_1)+S^{(2)}(B_2)\}.
\end{align}
The lack of additive property is crucial since this leads to the property that L-entropy does not vanish for the triangle state \eqref{eq:triangle}. Thus L-entropy does not measure the genuine tripartite entanglement in our criteria. 

\section{Black hole genuine multi-entropy curve}\label{sec:BHGMEC}

As an application of the multi-entropy to a black hole evaporation, the black hole multi-entropy curve was introduced and studied by \cite{Iizuka:2024pzm}. This is a natural generalization of the Page curve of entanglement entropy. 
The crucial point is that we approximate an evaporating black hole and its radiation with a Haar-random state for this purpose. This is based on the fact that through the gauge/gravity correspondence, asymptotic AdS quantum gravity is equivalent to ordinary quantum mechanical systems living on the boundary, and in general, such quantum systems are generally highly chaotic \cite{Sekino:2008he, Shenker:2013pqa} at the deconfinement phase dual to a black hole dominant phase \cite{Witten:1998qj, Witten:1998zw}. Since chaotic boundary states can be very well approximated by Haar-random states, it is reasonable to model the state of an evaporating black hole and its radiation using a boundary Haar-random state. In this section, similarly to the previous work,  we study the genuine multi-entropy ${\rm GM}^{(3)}(A:B:C)$ curve for the following tripartite systems: two Hawking radiation subsystems $A=\text{R1}$ and $B=\text{R2}$, and an evaporating black hole subsystem $C=\text{BH}$. 

More concretely, we first compute the black hole genuine $n=2$ R\'enyi tripartite entropy curve of ${\rm GM}_2^{(3)}(\text{R1}:\text{R2}:\text{BH})$ by using a single random tensor model as in the original work of Page curve \cite{Page:1993df, Page:1993wv}. The explicit expressions of R\'enyi multi-entropy are given by\footnote{As explained in footnote \ref{fn1}, our definition of $S^{(3)}_2$ differs from the definition in \cite{Iizuka:2024pzm} by a factor $1/2$.} \cite{Iizuka:2024pzm}
\begin{align}
 &   S^{(3)}_2(\text{R1}:\text{R2}:\text{BH})=-\frac{1}{2}\log \left[ \frac{ d_{\rm R}^2 d_{\rm BH}  (9 + 2 d_{\rm R}^2 + d_{\rm BH}^2 ) +  2 ( 2 d_{\rm R}^2 + d_{\rm BH}^2 + d_{\rm R}^4 + 2 d_{\rm R}^2 d_{\rm BH}^2 )}{(1 + d_{\rm BH} d_{\rm R}^2) (2 + d_{\rm BH} d_{\rm R}^2) (3 +  d_{\rm BH} d_{\rm R}^2)} \right],\label{MERS}\\
  &  \quad S^{(2)}_2(\text{R1})=S^{(2)}_2(\text{R2})=-\log\left[\frac{d_{\rm R}+d_{\rm R}d_{\rm BH}}{1 + d_{\rm BH} d_{\rm R}^2}\right], \; \quad S^{(2)}_2(\text{BH})=-\log\left[\frac{d_{\rm BH}+d_{\rm R}^2}{1 + d_{\rm BH} d_{\rm R}^2}\right],
\end{align}
where we set $\dim\mathcal{H}_\text{R1}=\dim\mathcal{H}_\text{R2}=d_\text{R}$ and $\dim\mathcal{H}_\text{BH}=d_\text{BH}$. We also fix the dimension of total system
\begin{align}
d_{\rm Total} = d_{\rm BH}d_{\rm R}^{2}  \mbox{ = fixed}.
\end{align}
By using $d_{\rm BH}=d_{\rm Total}/d_{\rm R}^{2}$, we can plot ${\rm GM}_2^{(3)}(\text{R1}:\text{R2}:\text{BH})$ as a function of $d_{\rm R}$.

\begin{figure}[t]
    \centering
    \includegraphics[width=10cm]{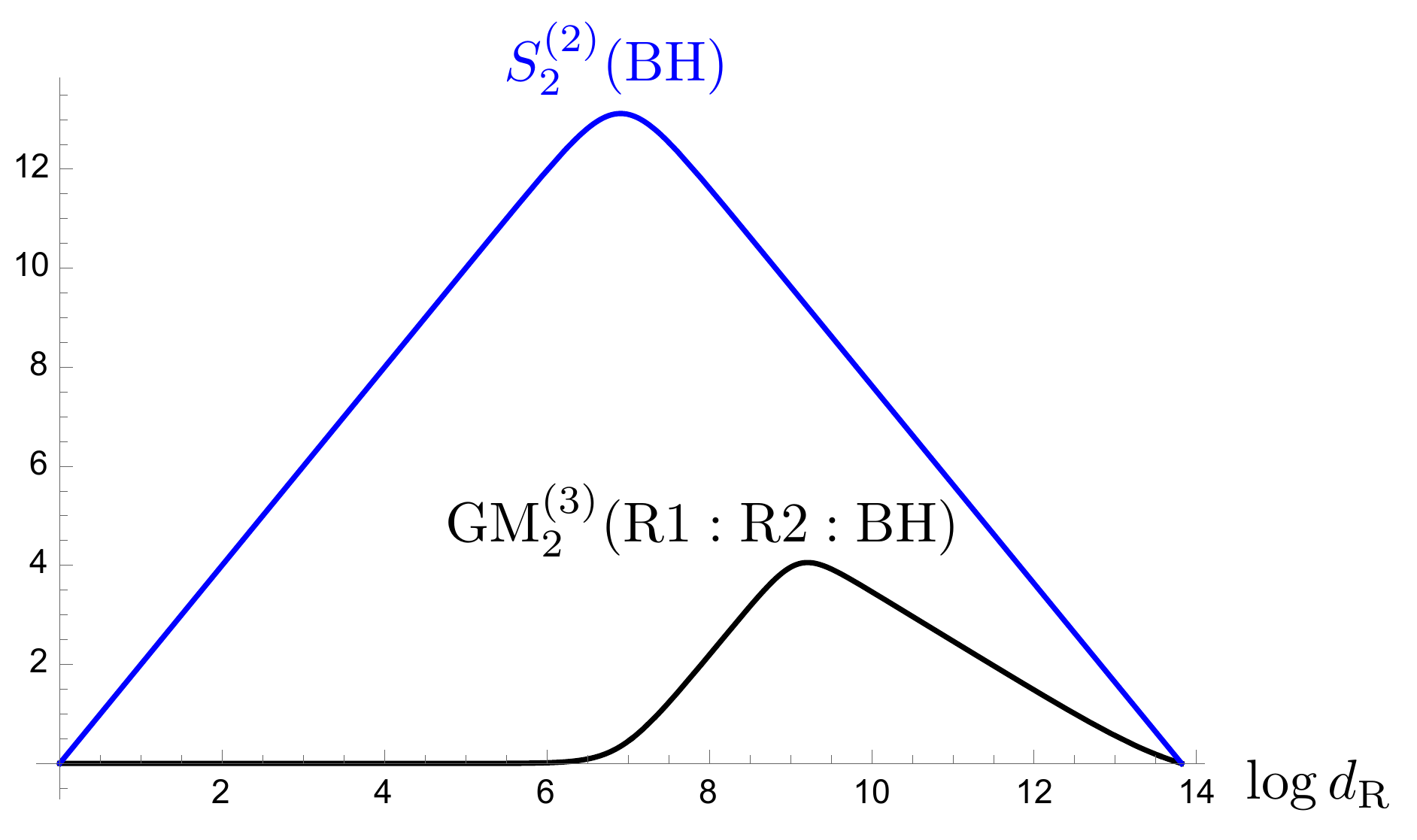}
    \caption{Black curve: black hole R\'enyi multi-entropy curve of ${\rm GM}_2^{(3)}(\text{R1}:\text{R2}:\text{BH})$. Blue curve: Page curve of $S^{(2)}_2(\text{BH})$. We fix the total dimension $d_{\rm Total} = d_{\rm BH}d_{\rm R}^{2}=10^{12}$.}
    \label{fig:GMEn2Curve}
\end{figure}

Figure \ref{fig:GMEn2Curve} shows the black hole genuine $\mathtt{q}=3, n=2$ R\'enyi multi-entropy curve of ${\rm GM}_2^{(3)}(\text{R1}:\text{R2}:\text{BH})$. 
The black curve shows ${\rm GM}_2^{(3)}(\text{R1}:\text{R2}:\text{BH})$ with $d_{\rm Total}=10^{12}$, where the horizontal axis is $\log d_\text{R}$.  
In the black hole $\mathtt{q}=3$ R\'enyi multi-entropy curve, there are only  one time scale, which is the multi-entropy time
\begin{align}
d_{\rm R} = d_{\rm BH} = d_{\rm Total}/d_{\rm R}^{2} \quad &\Leftrightarrow \quad d_{\rm R} = \left( d_{\rm Total} \right)^{1/3} \quad \mbox{(Multi-entropy time)}
\label{multiStime}.
\end{align}
For bipartite entanglement entropy, there is only one time scale, which is the Page time
\begin{align}
\label{Pagetime}
d_{\rm R}^{2} = d_{\rm BH} = d_{\rm Total}/d_{\rm R}^{2} \quad &\Leftrightarrow  \quad d_{\rm R} = \left( d_{\rm Total} \right)^{1/4} \quad \mbox{(Page time)}
\end{align}
Since $\mathtt{q}=3$ genuine multi-entropy is a linear combination of $\mathtt{q}=3$ and $\mathtt{q}=2$ multi-entropies,  there are only two time scales for that, and one can see this in Fig.~\ref{fig:GMEn2Curve}. 

The black hole genuine $n=2$ R\'enyi multi-entropy curve of ${\rm GM}_2^{(3)}(\text{R1}:\text{R2}:\text{BH})$ is initially zero and starts to increase after the Page time. Then, ${\rm GM}_2^{(3)}(\text{R1}:\text{R2}:\text{BH})$ starts to decrease from the multi-entropy time, and finally vanishes. In particular, we can see that the genuine $n=2$ R\'enyi multi-entropy ${\rm GM}_2^{(3)}(\text{R1}:\text{R2}:\text{BH})$ becomes nonzero and large {\it only after} the Page time. 
Thus, the Page time determines the time scale when it deviates from zero, and the multi-entropy time determines the time scale when it reaches maximum.

For the comparison of the magnitudes, we also plot the Page curve of $S^{(2)}_2(\text{BH})$ by the blue curve, which initially increases and starts to decrease from the Page time. One can see that even though the genuine tripartite multi-entropy contribution is smaller than the bipartite entanglement entropy, their magnitudes are similar in order. This implies that the genuine multi-partite entanglement is not small.

\begin{figure}[t]
    \centering
    \includegraphics[width=10cm]{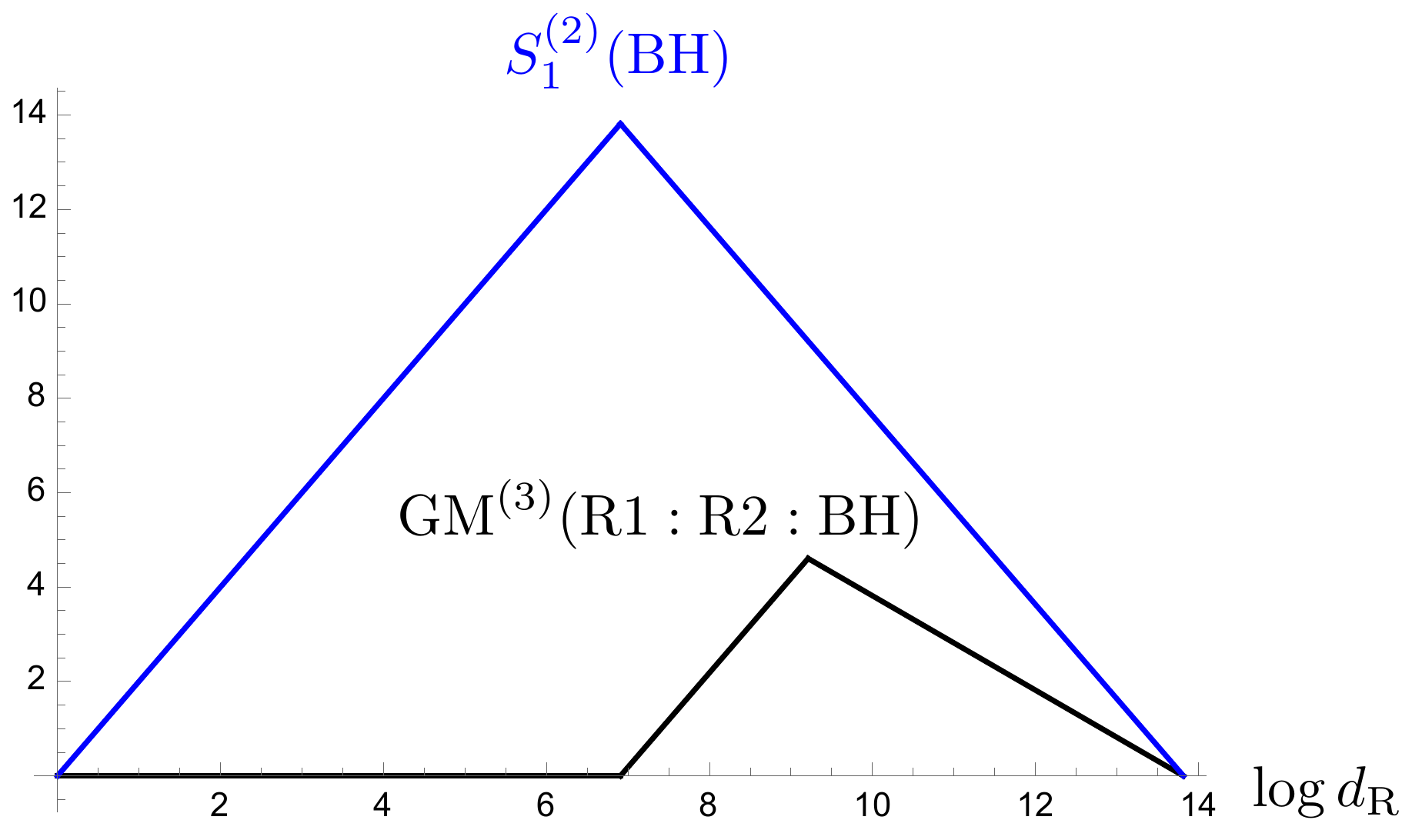}
    \caption{Asymptotic behaviors of ${\rm GM}^{(3)}(\text{R1}:\text{R2}:\text{BH})$ and $S^{(2)}_{n\to1}(\text{BH})$.}
    \label{fig:GMEn1Curve}
\end{figure}

Next, we study $n\to 1$ limit. 
The asymptotic behaviors of $S^{(3)}_n(\text{R1}:\text{R2}:\text{BH})$ at $d_\text{R}\ll d_\text{BH}$ and $d_\text{R}\gg d_\text{BH}$ was studied by \cite{Iizuka:2024pzm}. By using these expressions, we can plot the asymptotic behaviors in the black hole genuine multi-entropy curve of ${\rm GM}^{(3)}(\text{R1}:\text{R2}:\text{BH})$ at $n\to1$. Figure \ref{fig:GMEn1Curve} shows the asymptotic behaviors of ${\rm GM}^{(3)}(\text{R1}:\text{R2}:\text{BH})$ and $S^{(2)}_1(\text{BH})$. Since these are only the asymptotic behaviors, these plots are not smooth at the Page time and the multi-entropy time. Comparing Figures \ref{fig:GMEn2Curve} and \ref{fig:GMEn1Curve}, one can see that the qualitative behavior of ${\rm GM}_n^{(3)}(\text{R1}:\text{R2}:\text{BH})$ is independent of $n$. This is due to the fact that the asymptotic behaviors (5.23) in \cite{Iizuka:2024pzm} is not sensitive to $n$ if we divide them by a factor $n^{\mathtt{q}-2}$ as explained in footnote \ref{fn1}.  In particular, the two time scales, the Page time and the multi-entropy time, do not depend on $n$.

We also plot a similar curve of the Markov gap by using the study of reflected entropy in a single random tensor model by \cite{Akers:2021pvd}. R\'enyi generalization of Markov gap $h^{m,n}(A:B)$ is defined by \cite{Sohal:2023hst, Berthiere:2023gkx, Liu:2024ulq}  
\begin{align}
h^{m,n}(A:B):=S_{R}^{(m,n)}(A:B)-I^{(n)}(A:B),
\end{align}
where $I^{(n)}(A:B):=S_n^{(2)}(A)+S_n^{(2)}(B)-S_n^{(2)}(AB)$ is R\'enyi mutual information, and $S_{R}^{(m,n)}(A:B)$ is R\'enyi reflected entropy defined in \cite{Dutta:2019gen} as 
\begin{align}
S_{R}^{(m,n)}(A:B)&:=\frac{1}{1-n}\log \frac{Z_{m,n}}{(Z_{m,1})^n},\\
Z_{m,n}&:=\Tr_{AA^*}\left(\Tr_{BB^*}|\rho_{AB}^{m/2}\rangle\langle\rho_{AB}^{m/2}|\right)^n,
\end{align}
where $|\rho_{AB}^{m/2}\rangle$ is the canonical purified state of $\rho_{AB}^m$. Markov gap $h$ is defined by
\begin{align}
h:=\lim_{m\to1}\lim_{n\to1}h^{m,n}(A:B).
\end{align}
Here, the reduced density matrix $\rho_{AB}$ is defined by 
\begin{align}
\rho_{AB}:=\text{Tr}_C|\psi\rangle\langle\psi|,
\end{align}
for a given pure state $|\psi\rangle$ on $\mathcal{H}_A\otimes \mathcal{H}_B\otimes \mathcal{H}_C$. The explicit expression of $S_{R}^{(2,2)}(\text{R1}:\text{R2})$ for a random state $|\psi\rangle$ on $\mathcal{H}_\text{R1}\otimes \mathcal{H}_\text{R2}\otimes \mathcal{H}_\text{BH}$ is \cite{Iizuka:2024pzm}
\begin{align}
    S_{R}^{(2,2)}(\text{R1}:\text{R2})=-\log  \frac{ d_{\rm R}^2 d_{\rm BH}  (9 + 2 d_{\rm R}^2 + d_{\rm BH}^2 ) +  2 ( 2 d_{\rm R}^2 + d_{\rm BH}^2 + d_{\rm R}^4 + 2 d_{\rm R}^2 d_{\rm BH}^2 )}{d_\text{BH}^3d_\text{R}^2+d_\text{BH}^2(2d_\text{R}^4+4)+d_\text{BH}(d_\text{R}^6+10d_\text{R}^2)+4d_\text{R}^4+2},\label{RERS}
\end{align}
where we set $\dim\mathcal{H}_\text{R1}=\dim\mathcal{H}_\text{R2}=d_\text{R}$ and $\dim\mathcal{H}_\text{BH}=d_\text{BH}$.

\begin{figure}[t]
    \centering
    \includegraphics[width=10cm]{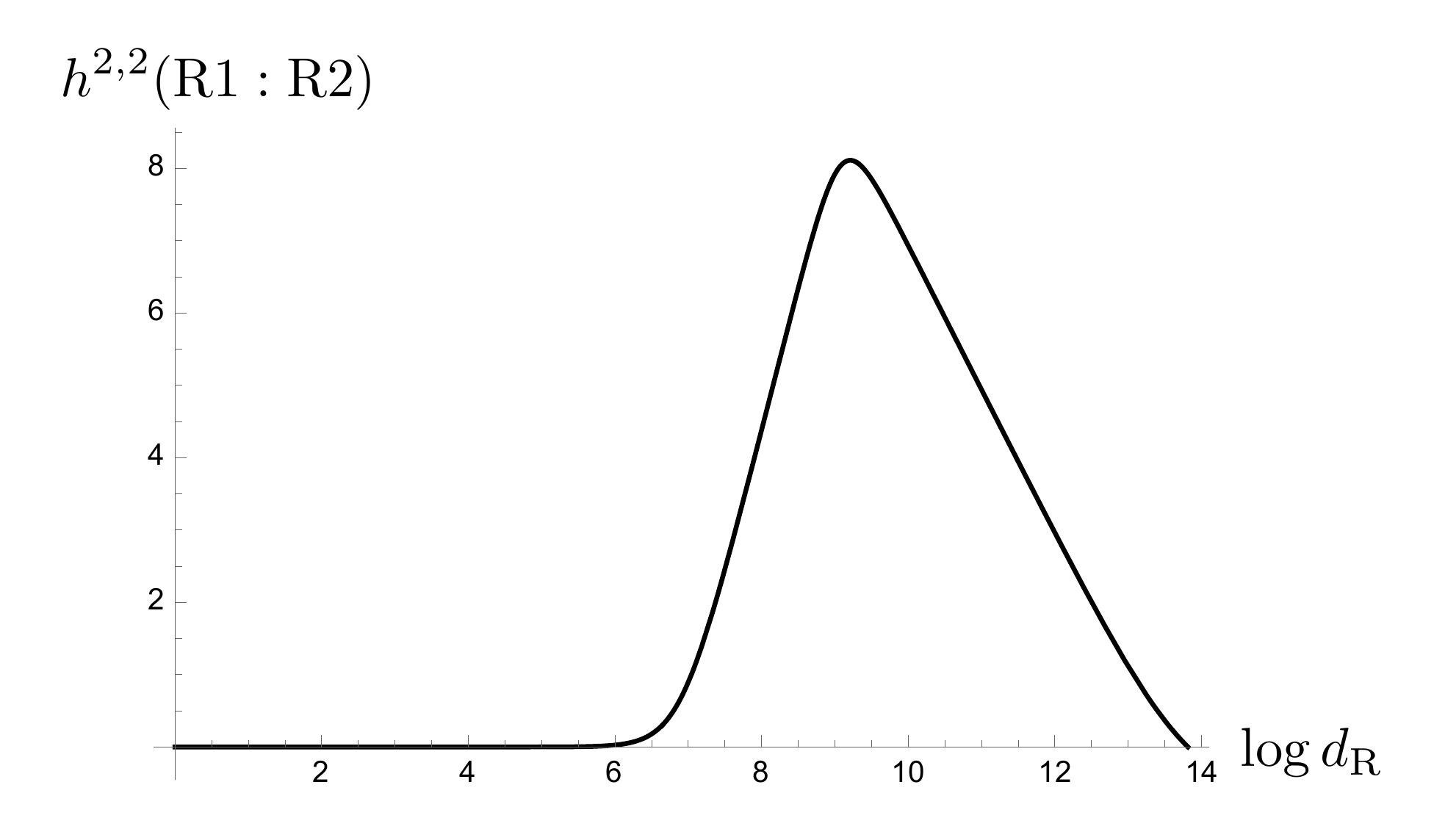}
    \caption{Curve of $h^{2,2}(\text{R1}:\text{R2})$ with the total dimension $d_{\rm Total} = d_{\rm BH}d_{\rm R}^{2}=10^{12}$.}
    \label{fig:MGm2n2Curve}
\end{figure}

\begin{figure}[t]
    \centering
    \includegraphics[width=10cm]{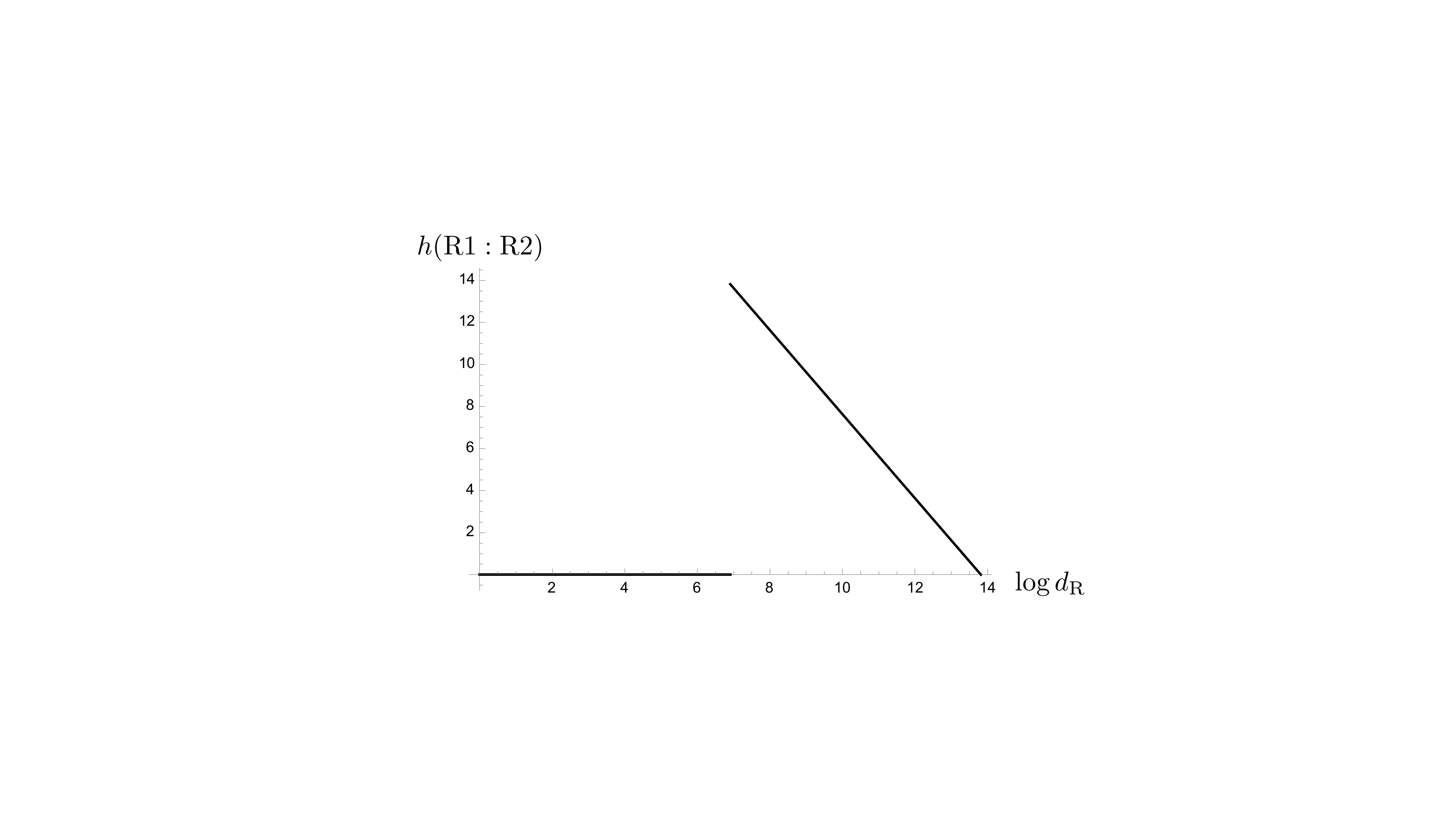}
    \caption{Asymptotic behavior of Markov gap $h(\text{R1}:\text{R2})$. There is the transition of asymptotic behavior at the Page time.}
    \label{fig:MGCurve}
\end{figure}

Figure \ref{fig:MGm2n2Curve} shows a curve of $h^{2,2}(\text{R1}:\text{R2})$ with the fixed total dimension $d_{\rm Total} = d_{\rm BH}d_{\rm R}^{2}=10^{12}$. This plot has a similar behavior to ${\rm GM}_2^{(3)}(\text{R1}:\text{R2}:\text{BH})$ in Figure \ref{fig:GMEn2Curve}. In fact, there is a relation\footnote{To derive the results in a single random tensor model such as eqs.~(\ref{MERS}) and (\ref{RERS}), an approximation that is valid when $d_{\rm Total} = d_{\rm BH}d_{\rm R}^{2}\gg1$ was used. } between ${\rm GM}_2^{(3)}(A:B:C)$ and $h^{2,2}(A:B)$ \cite{Liu:2024ulq}
\begin{align}
    {\rm GM}_2^{(3)}(A:B:C)=\frac{1}{2} \, {h^{2,2}(A:B)}.\label{RelationGMEMG}
\end{align}
One difference between ${\rm GM}_n^{(3)}(\text{R1}:\text{R2}:\text{BH})$ and $h^{m,n}(\text{R1}:\text{R2})$ is that the time scale of $h^{m,n}(\text{R1}:\text{R2})$ to start decreasing depends on $n$. From the result (3.96) in \cite{Akers:2021pvd}, this time scale of $h^{1,n}(\text{R1}:\text{R2})$ is given by
\begin{align}
d_{\rm R}^{2/n} = d_{\rm BH} = d_{\rm Total}/d_{\rm R}^{2} \quad &\Leftrightarrow  \quad d_{\rm R} = \left( d_{\rm Total} \right)^{\frac{n}{2(1+n)}},
\end{align}
which coincides the Page time when $n\to1$. Therefore, there is the transition in the asymptotic behavior of the Markov gap $h(\text{R1}:\text{R2})$ at the Page time as shown in Figure \ref{fig:MGCurve}.

Before we end, we summarize the characteristic natures of these curves. 
\begin{enumerate}
\item It is always only after the Page time, this genuine multi-entropy becomes large. 
This means before the Page time, essentially there is no contribution by genuine tripartite multi-entropy, compared with the bipartite entanglement.
\item As Fig.~\ref{fig:GMEn2Curve} and Fig.~\ref{fig:GMEn1Curve} show, after the Page time, the amount of genuine tripartite multi-entropy is not parametrically small compared with the bipartite entanglement. 
\item The genuine $\mathtt{q}=3$ multi-entropy curve Fig.~\ref{fig:GMEn1Curve} shows the peak at the multi-entropy time. On the other hand, the Markov gap Fig.~\ref{fig:MGCurve} shows the peak at the Page time. Even though both capture the genuine tripartite entanglement, the peak is at different time scales. This is because they are different measures. Note that a multi-entropy treats all subsystems on equal footing, however, the Markov gap does not. However, both quantities show nonzero value only after the Page time. 
\end{enumerate}

Quite interestingly,  the L-entropy \cite{Basak:2024uwc} also shows a very similar curve. See their Figure 17. 
However, as we commented before, since L-entropy does not vanish for the triangle state, their contributions is not necessary for genuine tripartite entanglement. In fact, quite similar curve can be obtained from the products of logarithmic negativity, see Appendix \ref{App:TripartiteNegativity} for more detail. 
It is interesting to investigate the difference between L-entropy and genuine multi-partite quantities in detail and see why Figure 17 of \cite{Basak:2024uwc} and Figure \ref{fig:GMEn2Curve} are similar.

\section{Why bipartite entanglement is not enough for holography}
\label{sec:proof}

\begin{figure}[t]
    \centering
    \includegraphics[width=13cm]{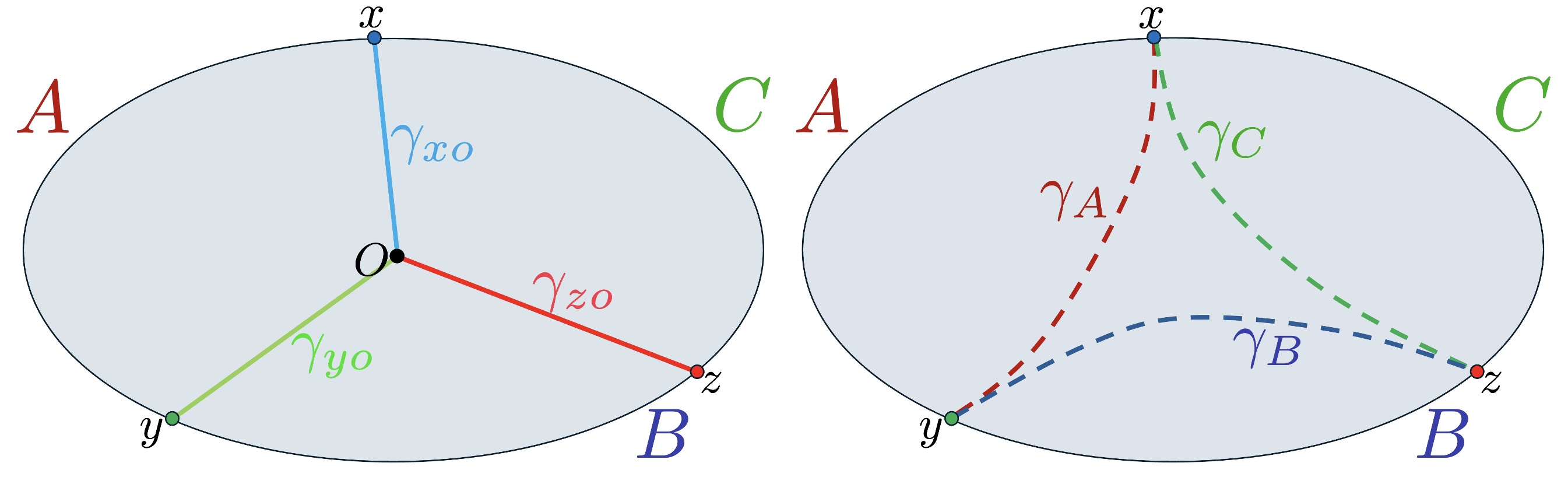}
    \caption{Left: Holographic multi-entropy is given by the sum of the three surfaces, $\gamma_{xo}$, $\gamma_{yo}$ and $\gamma_{zo}$. Right: RT surfaces $\gamma_A$, $\gamma_B$, $\gamma_C$ for the entanglement entropy $S_A$, $S_B$, $S_C$.}
    \label{fig:MERT}
\end{figure}

In this section, we show the inequality \eqref{inequality} for ${\rm GM}^{(3)}(A:B:C)$ holds in holographic settings\footnote{After writing the draft, we learned that the similar proof is also shown in \cite{Harper:2024ker} for $\mathtt{q}=3$ case.}. The proof is very simple. 
To show this, let's divide the boundary space into three regions/subsystems and let us call them $A$, $B$, $C$ respectively. 
The holographic multi-entropy $S^{(3)}(A:B:C)$ is given by the sum of the three surfaces, $\gamma_{xo}$, $\gamma_{yo}$ and $\gamma_{zo}$ \cite{Gadde:2022cqi, Gadde:2023zzj}, as 
\begin{align}
\label{holoSmultiE}
 S^{(3)}(A:B:C)  = \frac{1}{4 G_N} \left( \gamma_{xo} + \gamma_{yo} + \gamma_{zo} \right)
\end{align}
where $O$ is their meeting point. See the left figure of Fig.~\ref{fig:MERT}. In this section, we {\it assume}  that holographic multi-entropy is given by the formula as \eqref{holoSmultiE}. 

On the other hand, the entanglement entropy $S_A$, $S_B$, $S_C$ are given by the RT-surfaces $\gamma_A$, $\gamma_B$, $\gamma_C$ \cite{Ryu:2006bv}. 
Thus, 
\begin{align}
S_{A}+ S_{B}  +S_{C}  = \frac{1}{4 G_N} \left( \gamma_{A} + \gamma_{B} + \gamma_{C} \right)
\end{align}
See the right figure of Fig.~\ref{fig:MERT}, where RT-surfaces $\gamma_A$, $\gamma_B$, $\gamma_C$ represent the minimal surface connecting $x$ - $y$, $y$ - $z$ and $z$ - $x$ through the bulk, respectively.  

\begin{figure}[t]
    \centering
    \includegraphics[width=15cm]{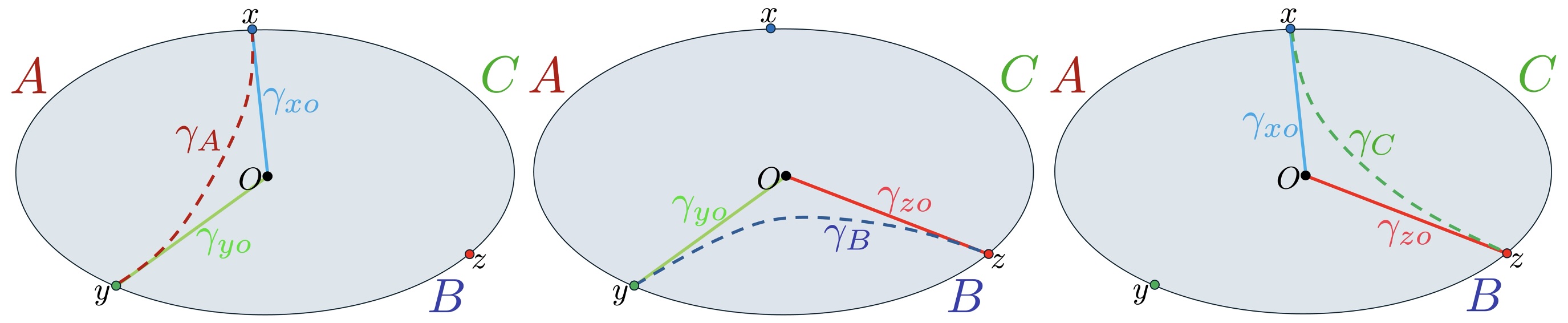}
    \caption{The genuine multi-entropy $\mathtt{q}=3$ reduces to the subtraction of 3 minimal surfaces. However, due to the fact that $\gamma_A$, $\gamma_B$, $\gamma_C$ are minimal surfaces, each parenthesis in \eqref{eachpare} is always positive quantities. Thus genuine multi-entropy is positive. }
    \label{fig:multi-rad}
\end{figure}

Thus, the genuine multi-entropy is always non-negative
\begin{align}
\label{eachpare}
& S^{(3)}(A:B:C)  - \frac{1}{2} \left( S_{A}+ S_{B}  +S_{C}  \right) \nonumber \\ 
& \,\, = \frac{1}{8 G_N} \left( \gamma_{xo} + \gamma_{yo} -  \gamma_{A}\right) +    \frac{1}{8 G_N} \left( \gamma_{yo} + \gamma_{zo} -  \gamma_{B}\right)  +  \frac{1}{8 G_N} \left( \gamma_{zo} + \gamma_{xo} -  \gamma_{C}\right)  > 0
\end{align}
since each parenthesis is always positive because RT-surfaces $\gamma_A$, $\gamma_B$, $\gamma_C$ are {\it minimal} surfaces. See Fig.~\ref{fig:multi-rad}. The similar proof is also shown in \cite{Harper:2024ker}. 

This holographic proof is analogous to the one given for the Markov gap given by \cite{Hayden:2021gno}. See their Fig.~3. In this way, even holographic-wise, genuine multi-entropy and the Markov gap share some similarities. However, there are also several differences. We summarize their analogies and differences here. 
\begin{enumerate}
\item Both the Markov gap and genuine multi-entropy are measures for genuine multi-partite entanglement in the sense that both quantities vanish for bipartite entangled states. 
\item However for GHZ state, which is one of the genuine multi-partite states, there is a difference. The Markov gap vanishes for the GHZ state, on the other hand, genuine multi-entropy gives a positive value for the GHZ state.  
\item Markov gap is defined for tripartite states in the sense that the reflected entropy is defined by tracing $C$ and leaving only subsystems $A$ and $B$. There is no way to furthermore divide the subsystem $C$ into finer subsystems. 
On the other hand, since the multi-entropy is defined in generic $\mathtt{q}$-partite system, it is possible to define a genuine multi-entropy even for $\mathtt{q} \ge 4$ case as well. We will comment on this in more detail later in the next section. 
\end{enumerate}
In this way, both the genuine multi-entropy and the Markov gap give evidence that in holography genuine tripartite entanglement is not small.

So far, we have focused on the $\mathtt{q}=3$ case. However, this proof can be easily generalized for higher $\mathtt{q}$ as well. The $n$-th Rényi multi-entropy of a pure state $\ket{\psi}$ on  $\mathtt{q}$-partite subsystems $A_1,A_2,\dots,A_\mathtt{q}$ is defined by \cite{Gadde:2022cqi, Gadde:2023zzj, Gadde:2023zni}
\begin{align}
S^{(\mathtt{q})}_n(A_1:A_2:\cdots:A_\mathtt{q}) &:= \frac{1}{1-n}\frac{1}{n^{\mathtt{q}-2}}\log \frac{Z^{(\mathtt{q})}_n}{(Z^{(\mathtt{q})}_1)^{n^{\mathtt{q}-1}}},\\
Z^{(\mathtt{q})}_n &:= \bra{\psi}^{\otimes n^{\mathtt{q}-1}} \Sigma_1(g_1)\Sigma_2(g_2)\cdots\Sigma_\mathtt{q}(g_\mathtt{q})\ket{\psi}^{\otimes n^{\mathtt{q}-1}},
\end{align}
where $\Sigma_k(g_k)$ are twist operators for the permutation action of \(g_k\) on indices of density matrices for $A_k$. The action of \(g_k\) can be expressed as
\begin{align}
\label{gkdefinition}
g_{k} & \cdot (x_1,\cdots,x_k,\cdots,x_{\mathtt{q}-1}) = (x_1,\cdots,x_{k}+1,\cdots,x_{\mathtt{q}-1}), \quad 1\le k \le \mathtt{q}-1, \\
g_\mathtt{q} &= e ,
\end{align}
where \((x_1,x_2,\cdots,x_{\mathtt{q}-1})\) represents an integer lattice point on a $(\mathtt{q}-1)$-dimensional hypercube of length \(n\), and we identify \(x_k= n + 1 \) with \(x_k=1\). The $\mathtt{q}$-partite multi-entropy $S^{(\mathtt{q})}(A_1:A_2:\cdots:A_\mathtt{q})$ is defined by taking the $n\to1$ limit as  
\begin{align}
S^{(\mathtt{q})}(A_1:A_2:\cdots:A_\mathtt{q}):=\lim_{n\to1}S^{(\mathtt{q})}_n(A_1:A_2:\cdots:A_\mathtt{q}).
\end{align}

As a $\mathtt{q}=4$ example, let us consider the following combination of $n\to 1$ limit, 
\begin{align}
\label{q4caseA}
S^{(4)}(A:B:C:D)-\frac{1}{2}\Bigl(&S^{(2)}(A:BCD)+S^{(2)}(B:CDA) \nonumber \\
 &+S^{(2)}(C:DAB) + S^{(2)}(D:ABC)\Bigr). 
\end{align}
Note that even though this combination excludes all of the bipartite contributions as ${\rm GM}_n^{(3)}(A:B:C)$ in \eqref{def_n_GM}, it contains {\it both} genuine tripartite and quadripartite entanglement. Since it includes genuine tripartite entanglement, we will not call this combination genuine quadripartite one. 

\begin{figure}[t]
    \centering
    \includegraphics[width=15cm]{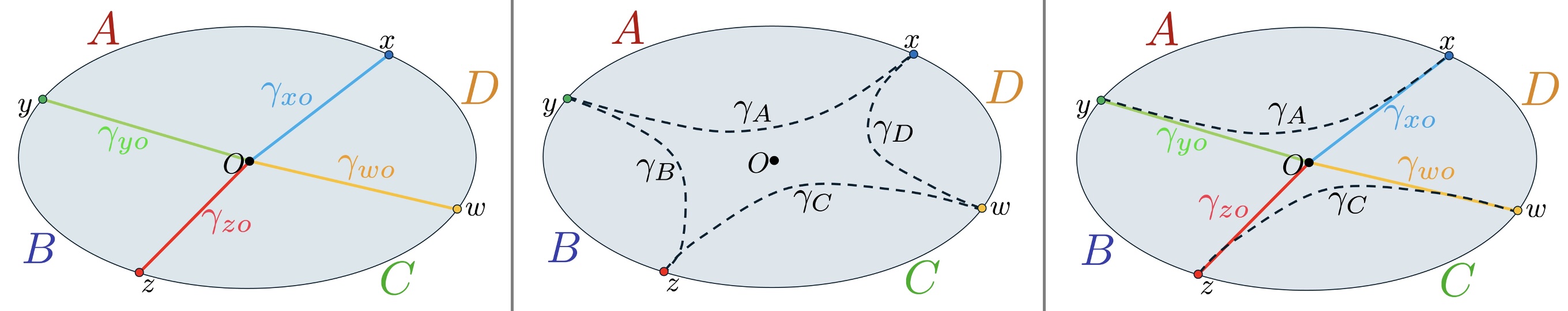}
    \caption{Left: For the case where holographic multi-entropy is given by the sum of the four surfaces, $\gamma_{xo}$, $\gamma_{yo}$, $\gamma_{zo}$ and $\gamma_{wo}$. Middle: RT surfaces $\gamma_A$, $\gamma_B$, $\gamma_C$, $\gamma_D$ for the entanglement entropy $S_A$, $S_B$, $S_C$, $S_D$. Right: \eqref{q4caseA} reduces to the summing over positive surfaces.}
    \label{fig:q41type}
\end{figure}

\begin{figure}[t]
    \centering
    \includegraphics[width=15cm]{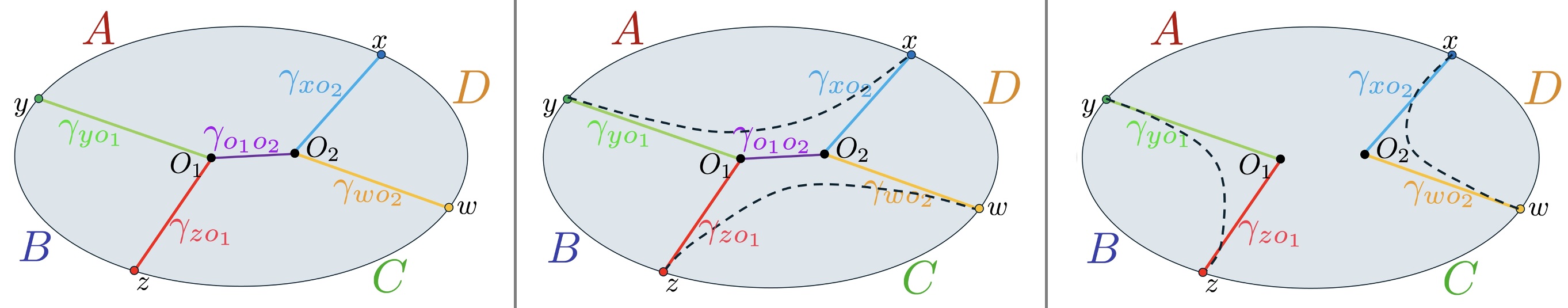}
    \caption{Left: For the case where holographic multi-entropy is given by the sum of the five surfaces, $\gamma_{xo_2}$, $\gamma_{yo_1}$, $\gamma_{zo_1}$, $\gamma_{wo_2}$ and $\gamma_{o_1o_2}$. Middle and Right: Subtracting the RT surfaces gives positive in both cases.}
    \label{fig:q42type}
\end{figure}

It is straightforward to show this combination is always positive in holography
\begin{align}
\label{q4case11}
S^{(4)}(A:B:C:D)&-\frac{1}{2}\Bigl(S^{(2)}(A:BCD)+S^{(2)}(B:CDA) \nonumber \\
&\qquad +S^{(2)}(C:DAB) + S^{(2)}(D:ABC)\Bigr)  > 0.\; 
\end{align}
The proof is the same as the previous section. See Fig,~\ref{fig:q41type} and ~\ref{fig:q42type}. Again we {\it assume}  that holographic multi-entropy is given by the formula as Left figures of Fig.~\ref{fig:q41type} and ~\ref{fig:q42type} depending on the subregions $A$, $B$, $C$, $D$. Crucial assumption here is that holographic dual admits ``intersection surfaces'', such as $O$ in Fig.~\ref{fig:q41type} and $O_1$, $O_2$ in Fig.~\ref{fig:q42type}. 

\begin{enumerate}
\item For the case where holographic multi-entropy is given by the sum of the four surfaces, $\gamma_{xo}$, $\gamma_{yo}$, $\gamma_{zo}$ and $\gamma_{wo}$ as Fig.~\ref{fig:q41type}, 
\begin{align}
\label{q4caseB}
&S^{(4)}(A:B:C:D)-\frac{1}{2}\Bigl(S^{(2)}(A:BCD)+S^{(2)}(B:CDA) \nonumber \\
&\qquad +S^{(2)}(C:DAB) + S^{(2)}(D:ABC)\Bigr) 
\\
& \,\, = \frac{1}{8 G_N} \left( \gamma_{xo} + \gamma_{yo} -  \gamma_{A}\right) +    \frac{1}{8 G_N} \left( \gamma_{yo} + \gamma_{zo} -  \gamma_{B}\right)  \nonumber \\ 
& \,\, \,\,\,\, +  \frac{1}{8 G_N} \left( \gamma_{zo} + \gamma_{wo} -  \gamma_{C}\right) +  \frac{1}{8 G_N} \left( \gamma_{wo} + \gamma_{xo} -  \gamma_{D}\right)  > 0.\;
\end{align}
\item For the case where holographic multi-entropy is given by the sum of the five surfaces, $\gamma_{xo_2}$, $\gamma_{yo_1}$, $\gamma_{zo_1}$, $\gamma_{wo_2}$ and $\gamma_{o_1 o_2}$ as Fig.~\ref{fig:q42type}, 
\begin{align}
\label{q4caseB}
&S^{(4)}(A:B:C:D)-\frac{1}{2}\Bigl(S^{(2)}(A:BCD)+S^{(2)}(B:CDA) \nonumber \\
&\qquad +S^{(2)}(C:DAB) + S^{(2)}(D:ABC)\Bigr) 
\\
& \,\, = \frac{1}{8 G_N} \left( \gamma_{xo_2} +  \gamma_{o_1 o_2} + \gamma_{yo_1}  -  \gamma_{A}\right) +    \frac{1}{8 G_N} \left( \gamma_{yo_1} + \gamma_{zo_1} -  \gamma_{B}\right)  \nonumber \\ 
& \,\, \,\,\,\, +  \frac{1}{8 G_N} \left( \gamma_{zo_1} +  \gamma_{o_1 o_2}+ \gamma_{wo_2} -  \gamma_{C}\right) +  \frac{1}{8 G_N} \left( \gamma_{wo_2} + \gamma_{xo_2} -  \gamma_{D}\right)  > 0.\;
\end{align}
\end{enumerate}
Therefore, in any case\footnote{In fact, the configuration in Fig.~\ref{fig:q41type} never become dominant.}, we obtain the inequality (\ref{q4case11}).

It is pretty straightforward to generalize this to general $\mathtt{q}$, where we obtain the following inequality 
\begin{align}
\label{Sqinequality}
S^{(\mathtt{q})}(A_1: A_2: \cdots :A_\mathtt{q})-\frac{1}{2}\left( \sum_{k=1}^\mathtt{q} S^{(2)}(A_k) \right)  > 0 
\end{align}
from holography, where $S^{(2)}(A_k)$ is bipartite entanglement entropy between subregion $A_k$ and the rest.
Here, we assume the bulk geometry is the dual of a pure state on the subsystems $A_1,A_2,\dots,A_\mathtt{q}$, and the boundary subregions are all connected.
Note that the left-hand side of \eqref{Sqinequality} excludes all bipartite entanglement contributions. Thus, this generic $\mathtt{q}$ inequality implies that in holography, bipartite entanglement is not enough. 


\begin{figure}[t]
    \centering
    \includegraphics[width=11cm]{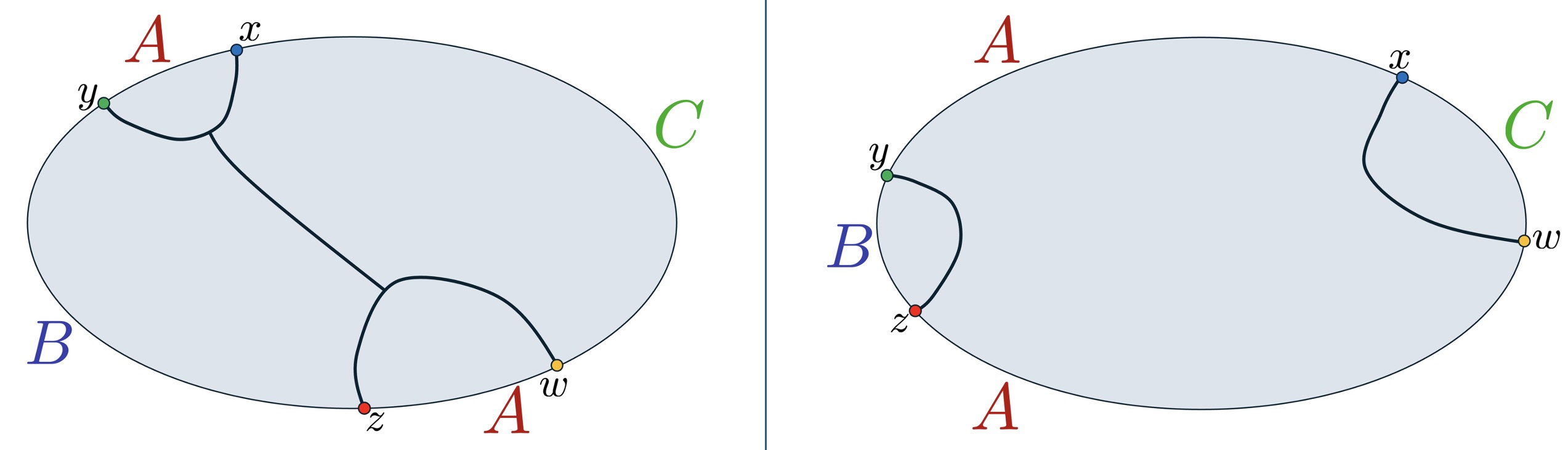}
    \caption{For the case where holographic multi-entropy is given by surfaces admitting ``intersection surfaces'' as the left figure, \eqref{Sqinequality2} is positive, but if it is not, as shown in the right figure, \eqref{Sqinequality2} is saturated.}
    \label{fig:saturation}
\end{figure}

So far we have assumed that the boundary subregions $A_1,A_2,\dots,A_\mathtt{q}$ are connected subregions. It is straightforward to generalize these proofs to the cases where the boundary subregions are disconnected\footnote{We thank Simon Lin for asking this point.}. However, for such cases, there is a possibility that in the bulk, holographic dual does not admit ``intersection surfaces'', such as $O$ in Fig.~\ref{fig:q41type} and $O_1$, $O_2$ in Fig.~\ref{fig:q42type}. For example, see figure 11 in \cite{Harper:2024ker}. 
In such cases, there is a case that an inequality \eqref{Sqinequality} can be saturated. For example, if the disconnected subregion $A$ is much larger than subregions $B$ and $C$ as shown in the right figure of Fig.~\ref{fig:saturation}, the holographic multi-entropy and the holographic entanglement entropy are given by
\begin{align}
S^{(3)}(A:B:C)=S_A=\frac{1}{4 G_N} \left( \gamma_{B} + \gamma_{C}\right), \;\;\; S_B=\frac{1}{4 G_N} \gamma_{B}, \;\;\; S_C=\frac{1}{4 G_N} \gamma_{C}.\label{HMEDisconnected}
\end{align} 
Thus, if there are ``intersection surfaces'' in the bulk for holographic dual of multi-entropy,  inequality \eqref{Sqinequality} holds, but if not, the inequality \eqref{Sqinequality} can be saturated. Thus, we conclude 
\begin{align}
\label{Sqinequality2}
S^{(\mathtt{q})}(A_1: A_2: \cdots :A_\mathtt{q})-\frac{1}{2}\left( \sum_{k=1}^\mathtt{q} S^{(2)}(A_k) \right)  \ge 0.  
\end{align}
The equality can hold only if there is no ``intersection surfaces'' in the bulk. 

This argument is parallel to the case of the holographic Markov gap\footnote{\label{posimarkovproof}See Figure 3 in \cite{Hayden:2021gno}. The holographic Markov gap is proportional to $\left( \mbox{KRT}(A) - \mbox{RT}(A) \right) +\left( \mbox{KRT}(B) - \mbox{RT}(B) \right)$. 
The non-negativity of the geometrical quantities (as indicated in each parenthesis) is parallel to our argument.}. 
In holography, there are cases where the Markov gap becomes zero as well.

\section{Generic $\mathtt{q}$ genuine multi-entropy}\label{sec:GMEq4}
\subsection{$\mathtt{q}=4$ genuine multi-entropy}
Similarly to $\mathtt{q}=3$, by taking appropriate linear combinations of multi-entropy, one can define genuine multi-entropy for the $\mathtt{q}\ge4$ case as well. In fact, for the $\mathtt{q}=4$ case, we can define the following linear combination; 
\begin{align}
\label{quadripartite1}
&{\rm GM}_n^{(4)}(A:B:C:D):=S_n^{(4)}(A:B:C:D)
-\frac{1}{3}\Big(S_n^{(3)}(AB:C:D)+S_n^{(3)}(AC:B:D) \notag\\
& \qquad \qquad +S_n^{(3)}(AD:B:C) +S_n^{(3)}(BC:A:D)+S_n^{(3)}(BD:A:C)+S_n^{(3)}(CD:A:B)\Big)\notag\\
&\qquad+a\left(S_n^{(2)}(AB:CD)+S_n^{(2)}(AC:BD)+S_n^{(2)}(AD:BC)\right)\notag\\
&\qquad +b\left(S_n^{(2)}(ABC:D)+S_n^{(2)}(ABD:C)+S_n^{(2)}(ACD:B)+S_n^{(2)}(BCD:A)\right), 
\end{align}
where $S_n^{(\mathtt{q})}$ is $\mathtt{q}$-partite R\'enyi multi-entropy, and $a$ and $b$ are parameters satisfying 
\be\label{const1}
 a+b = \frac{1}{3}  \,.
\ee
This can also be expressed in terms of the genuine tripartite multi-entropies such as ${\rm GM}_n^{(3)}(AB:C:D)$ for three subsystems $AB$, $C$, $D$ defined in \eqref{def_n_GM} as 
\begin{align}
\label{quadripartite2}
& {\rm GM}_n^{(4)}(A:B:C:D):=S_n^{(4)}(A:B:C:D) -\frac{1}{3}\Big( {\rm GM}_n^{(3)}(AB:C:D)+ {\rm GM}_n^{(3)}(AC:B:D)\notag \\ &\quad +{\rm GM}_n^{(3)}(AD:B:C)+{\rm GM}_n^{(3)}(BC:A:D)+{\rm GM}_n^{(3)}(BD:A:C)+{\rm GM}_n^{(3)}(CD:A:B)\Big)\notag\\
&\qquad +
\tilde{a}  \left(S_n^{(2)}(AB:CD)+S_n^{(2)}(AC:BD)+S_n^{(2)}(AD:BC)\right)\notag\\
&\qquad +
\tilde{b} \left(S_n^{(2)}(ABC:D)+S_n^{(2)}(ABD:C)+S_n^{(2)}(ACD:B)+S_n^{(2)}(BCD:A)\right),
\end{align}
where 
\begin{align}
\label{quadripartite3}
 \tilde{a} = a - \frac{1}{3}\,, \quad  \tilde{b} = b - \frac{1}{2}  \,, \quad \tilde{a} + \tilde{b} = - \frac{1}{2} \,. 
\end{align}
The claim is that this combination excludes the tripartite and bipartite entanglement and captures only the quadripartite entanglement only. This means, ${\rm GM}_n^{(4)}(A:B:C:D)$ is zero for pure states with tripartite or bipartite entanglement such as
\begin{align}
\label{examplestates}
|\psi_1\rangle_{ABC}\otimes |\psi_2\rangle_D, \;\;\; |\psi_1\rangle_{AB}\otimes |\psi_2\rangle_{CD} \,.
\end{align}
This can be seen as follows. 
\begin{enumerate}
\item[A:] Consider a tripartite state, for example, $|\psi_1\rangle_{ABC}\otimes |\psi_2\rangle_D$. In this case, we do not need to contract the index of subsystem $D$, and the contraction of $n^3$ reduced density matrices for $S_n^{(4)}$ is factorized to $n$ product of the contraction of $n^2$ reduced density matrices due to the definition of the multi-entropy.
Then, the following identities hold
\begin{align}
\label{id1}
S_n^{(4)}(A:B:C:D)\qquad &\nonumber \\
=S_n^{(3)}(AD:B:C) &=S_n^{(3)}(BD:A:C)=S_n^{(3)}(CD:A:B),\\
\label{id2}
S_n^{(3)}(AB:C:D)&=S_n^{(2)}(AB:CD)=S_n^{(2)}(ABD:C),\\
\label{id3}
S_n^{(3)}(AC:B:D)&=S_n^{(2)}(AC:BD)=S_n^{(2)}(ACD:B),\\
\label{id4}
S_n^{(3)}(BC:A:D)&=S_n^{(2)}(AD:BC) =S_n^{(2)}(BCD:A),\\
\label{id5}
& S_n^{(2)}(ABC:D)=0.
\end{align}
\eqref{id1} - \eqref{id4} give 
\begin{align}
S_n^{(4)}(A:B:C:D) \qquad \qquad \qquad \qquad \qquad \qquad \nonumber  \qquad \qquad \qquad \qquad & \\
\quad -\frac{1}{3}\left(S_n^{(3)}(AD:B:C)+S_n^{(3)}(BD:A:C)+S_n^{(3)}(CD:A:B)\right)&=0,\\
-\frac{1}{3}S_n^{(3)}(AB:C:D)+a S_n^{(2)}(AB:CD)+b S_n^{(2)}(ABD:C)&=0,\\
-\frac{1}{3}S_n^{(3)}(AC:B:D)+a S_n^{(2)}(AC:BD)+b S_n^{(2)}(ACD:B)&=0,\\
-\frac{1}{3}S_n^{(3)}(BC:A:D)+a S_n^{(2)}(AD:BC)+b S_n^{(2)}(BCD:A)&=0.
\end{align}
Combined these with \eqref{id5}, one can see that ${\rm GM}_n^{(4)}(A:B:C:D)$ given by \eqref{quadripartite1} vanishes.  
\item[B:] Similarly, consider a bipartite state, for example, $ |\psi_1\rangle_{AB}\otimes |\psi_2\rangle_{CD}$. Its  contribution is decomposed into the sum of contributions of  $|\psi_1\rangle_{AB}$ and $|\psi_2\rangle_{CD}$ 
due to the additive property (\ref{additive}). Thus, we consider only for a state $|\psi_1\rangle_{AB}$, where subsystems $C$ and $D$ are not entangled. Note that in this case, since genuine tripartite multi-entropies such as ${\rm GM}_n^{(3)}(AC:B:D)$ do not contain any bipartite contributions, they vanish. In this case, the following identities hold 
\begin{align}
\label{id6}
S_n^{(4)}(A:B:C:D) &=S_n^{(2)}(AC:BD) =S_n^{(2)}(AD:BC) \nonumber \\
&=S_n^{(2)}(ACD:B) =S_n^{(2)}(BCD:A),\\
\label{id7}
 S_n^{(2)}(AB:CD) &= S_n^{(2)}(ABC:D)= S_n^{(2)}(ABD:C) =  0 \,.
\end{align}
\eqref{id6} can be written as 
\begin{align}
S_n^{(4)}(A:B:C:D) &+ \tilde{a} \left( S_n^{(2)}(AC:BD) + S_n^{(2)}(AD:BC) \right) \nonumber \\
&+ \tilde{b} \left( S_n^{(2)}(ACD:B) + S_n^{(2)}(BCD:A) \right)  = 0
\end{align}
Using this with \eqref{id7}, one can see that ${\rm GM}_n^{(4)}(A:B:C:D)$ given by 
\eqref{quadripartite2} vanishes. 
\end{enumerate}

Thus, ${\rm GM}_n^{(4)}(A:B:C:D)$ for tensor products of states like \eqref{examplestates} are zero. 

\subsection{$\mathtt{q}=4$ genuine multi-entropy curve}
\begin{figure}[t]
  \centering
  \begin{subfigure}[t]{.32\textwidth}
    \centering
    \includegraphics[width=\linewidth]{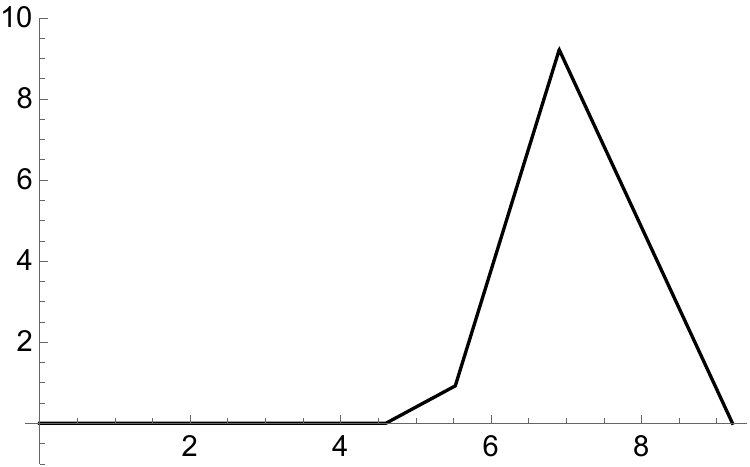}
        \put(-20,-10){$\log d_{\text{R}}$}
    \put(-150,90){${\rm GM}^{(4)}$}
    \caption{$a=1/2,b=-1/6$}
  \end{subfigure}
  \begin{subfigure}[t]{.32\textwidth}
    \centering
    \includegraphics[width=\linewidth]{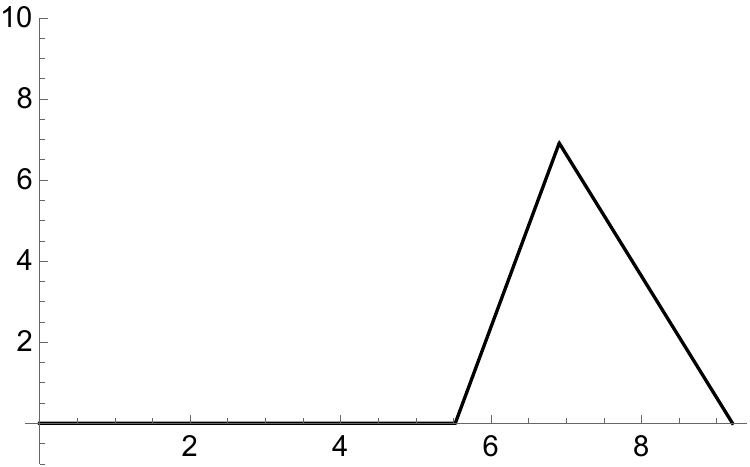}
    \put(-20,-10){$\log d_{\text{R}}$}
    \put(-150,90){${\rm GM}^{(4)}$}
    \caption{$a=1/3,b=0$}
  \end{subfigure}
  \begin{subfigure}[t]{.32\textwidth}
    \centering
    \includegraphics[width=\linewidth]{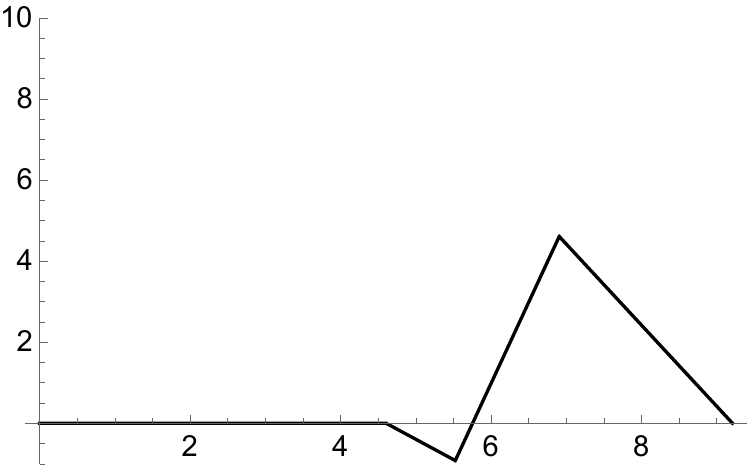}
    \put(-20,-10){$\log d_{\text{R}}$}
    \put(-150,90){${\rm GM}^{(4)}$}
    \caption{$a=1/6,b=1/6$}
  \end{subfigure}
  \caption{Asymptotic behaviors of ${\rm GM}^{(4)}(\text{R1}:\text{R2}:\text{R3}:\text{BH})$ for a Haar-random state with different values of $a$ and $b$, where $\dim\mathcal{H}_\text{R1}=\dim\mathcal{H}_\text{R2}=\dim\mathcal{H}_\text{R3}=d_\text{R}$, $\dim\mathcal{H}_\text{BH}=d_\text{BH}$, and we fix $d_{\rm Total} = d_{\rm BH}d_{\rm R}^{3}=10^{12}$. There are three time scales: (1) $d_\text{R}^3=d_\text{BH}$, (2) $d_\text{R}^2=d_\text{BH}$, and (3) $d_\text{R}=d_\text{BH}$.}\label{fig:GM4}
\end{figure}
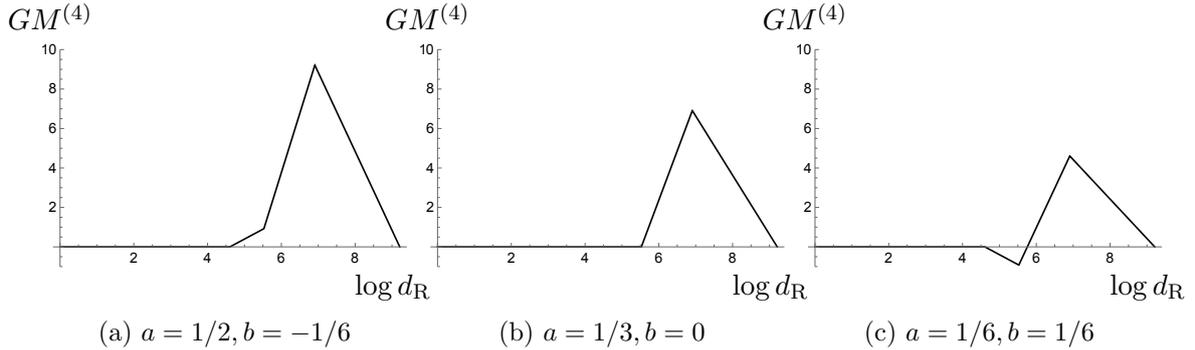

As a generalization of Fig.~\ref{fig:GMEn1Curve}, we plot the black hole genuine multi-entropy curve of ${\rm GM}^{(4)}(\text{R1}:\text{R2}:\text{R3}:\text{BH})$ in Fig.~\ref{fig:GM4}, where $\dim\mathcal{H}_\text{R1}=\dim\mathcal{H}_\text{R2}=\dim\mathcal{H}_\text{R3}=d_\text{R}$, $\dim\mathcal{H}_\text{BH}=d_\text{BH}$, and we fix $d_{\rm Total} = d_{\rm BH}d_{\rm R}^{3}=10^{12}$. One can see that ${\rm GM}^{(4)}$ can be negative if $b$ is positive. Thus, if we impose that ${\rm GM}^{(4)}$ is non-negative, $b$ should be non-positive. The important point we emphasize is that, as we have seen for the $\mathtt{q}=3$ case, $\mathtt{q}=4$ genuine multi-entropy is almost zero until the Page time, where $d_\text{R}^3=d_\text{BH}$. They become important only after the Page time for whatever value of the parameter.

For comparison, in Appendix~\ref{App:CforIq4}, we present a detailed evaluation of the $\mathtt{q}$-partite information $I_{\mathtt{q}}$ during black hole evaporation. For $\mathtt{q}=3$, $I_3$ fails to reflect the emergence of multipartite entanglement. In contrast, for $\mathtt{q}=4$, the behavior of $I_{\mathtt{q}}$—up to an overall sign—exhibits a marked increase only after the Page time, as illustrated in Fig.~\ref{fig:App}. 

\subsection{How to construct generic $\mathtt{q}$ genuine multi-entropy}
It is straightforward to generalize this construction for $\mathtt{q}\ge5$. General prescription to construct the genuine $\mathtt{q}$-partite multi-entropy is followings. 
\begin{enumerate}
\item First, consider a general expression for the linear combination of $S_n^{(\mathtt{q})}$, $S_n^{(\mathtt{q}-1)}$, $\dots$, $S_n^{(2)}$ with symmetry in all the parties. 
\item Then, determine each coefficient of the linear combination so that the genuine $\mathtt{q}$-partite multi-entropy vanishes for states with $(\mathtt{q}-1)$-partite entanglement such as
\begin{align}
|\psi\rangle=|\psi_1\rangle_{A_1A_2\dots A_{\mathtt{q}-1}}\otimes |\psi_2\rangle_{A_\mathtt{q}}.
\end{align}
\end{enumerate}
We emphasize that the symmetric property (\ref{symmetric}) and the additive property (\ref{additive}) are crucial in this construction.
It is interesting to investigate these higher $\mathtt{q}$ genuine multi-partite entanglement in more detail from both CFTs and also holography. It is quite interesting to investigate 
more details for higher $\mathtt{q}$ case. 

\section{Discussions and open questions} 
\label{sec:discussion}
The central message of this paper is that one can define a genuine multi-entropy for any $\mathtt{q} \ge 3$. Even though a multi-entropy $S^{(\mathtt{q})}$ proposed in \cite{Gadde:2022cqi,Penington:2022dhr,Gadde:2023zzj} is an interesting quantity that captures multi-partite entanglement, it is not particularly useful on its own, since it also receives contributions from lower-partite $\tilde{\mathtt{q}} < \mathtt{q}$ entanglement. 
In other words, $S^{(\mathtt{q})}$ can be nonzero even in the presence of only $\tilde{\mathtt{q}}$-partite entanglement. 

To isolate the contribution from genuine $\mathtt{q}$-partite entanglement, we define the genuine multi-entropy, ${\rm GM}^{(\mathtt{q})}$, which vanishes unless genuine $\mathtt{q}$-partite entanglement is present. Intuitively, one can think of ${\rm GM}^{(\mathtt{q})}$ as an “irreducible representation decomposition”, 
 separating out the distinct contributions from different levels of multipartite entanglement contained within $S^{(\mathtt{q})}$.

The remainder of this paper can be summarized as follows:
\begin{enumerate}
\item \underline{Explicit construction of genuine multi-entropies}: \\
We construct explicitly $\mathtt{q}=3$ and $\mathtt{q}=4$ examples and give a concise yet general prescription on how to construct generic $\mathtt{q}$ genuine multi-entropy for $\mathtt{q} \ge 5$ cases.  
\item \underline{Application to evaporating black holes}: \\
Applying our framework to the case of black hole evaporation, we compute the evolution of ${\rm GM}^{(q)}$ as a function of time  for both $\mathtt{q}=3$ and $\mathtt{q}=4$ cases, extending the analysis of \cite{Iizuka:2024pzm}. In this setup, the Hawking radiation is partitioned into $\mathtt{q}-1$ subsystems. We find that genuine multipartite entanglement becomes significant only after the Page time; before that, bipartite entanglement between the black hole and the radiation suffices to characterize the system.
\item \underline{Application to holography}: \\ 
We prove that the non-bipartite components of the multi-entropy are always positive and of order $\order{1/G_N}$, as long as the boundary subregions are connected. This implies that genuine multipartite entanglement is non-negligible in holographic systems and plays an important role in their entanglement structure.
\end{enumerate}

We conclude this paper with a discussion of open questions and future directions. 
\begin{enumerate}
\item We conjectured that $c_W>1$ in (\ref{Wconjecture}) by evaluating $S^{(3)}_n$ of the W-state for small $n$. Can we rigorously prove $c_W>1$? And, in three qubit systems, what state gives the maximum value of multi-entropy and genuine tripartite multi-entropy?
\item By assuming the holographic prescription of multi-entropy \cite{Gadde:2022cqi, Gadde:2023zzj}, we argued the inequality \eqref{Sqinequality2}. Does this inequality hold even for non-holographic CFTs or general spin systems? Or does it only work in special cases, like the monogamy of holographic mutual information \cite{Hayden:2011ag}?
\item We assumed the holographic prescription of multi-entropy at $n\to1$ \cite{Gadde:2022cqi, Gadde:2023zzj}. However, as demonstrated by \cite{Penington:2022dhr}, in holographic CFTs, $S^{(3)}_n$ with $n\ge3$ is not dual to the minimal triway cuts in the bulk, and thus the validity of holographic assumption at $n\to1$ is unclear. At least, for pure states in RTN, the holographic assumption is plausible \cite{Penington:2022dhr}, and our geometrical proof could be applied. 
\item Further detailed analysis on the genuine multi-entropy for $\mathtt{q} \ge 4$ is needed for both holography and generic quantum states. We would like to report it in the near future.
\end{enumerate}

\acknowledgments
We would like to thank Simon Lin for the collaboration on \cite{Iizuka:2024pzm} and also for the helpful comments on the draft and discussion on the Markov gap. 
The work of N.I. was supported in part by JSPS KAKENHI Grant Number 18K03619, MEXT KAKENHI Grant-in-Aid for Transformative Research Areas A “Extreme Universe” No. 21H05184. 
\appendix

\appendix

\section{The maximum value of $S^{(3)}_{2}$ in W-class}\label{App:MaxMEW}
Let us consider a pure state in W-class \cite{Dur:2000zz}
\begin{align}
|\psi_W\rangle=\sqrt{a}|001\rangle+\sqrt{b}|010\rangle+\sqrt{c}|100\rangle+\sqrt{d}|000\rangle,
\end{align}
where $a,b,c>0$ and $d=1-a-b-c\ge0$. We want to find the value of $a,b,c,d$ to maximize $S^{(3)}_{2 \; |\psi_W\rangle}$. To maximize $S^{(3)}_{2}$, replica partition function $Z_2^{(3)}$ should be minimum. By using $c=1-a-b-d$, one can express $Z_2^{(3)}$ of $|\psi_W\rangle$ as
\begin{align}
Z_2^{(3)}&=\left(2 a^2+2 a (b-1)+2 b^2-2 b+1\right)^2\notag\\
&+d(4 (2 a^3 + 2 a^2 (-1 + b) + a (1 - 2 b + 2 b^2) + b (1 - 2 b + 2 b^2)))+(4 a^2 + 4 b^2) d^2\notag\\
&=:f(a,b,d).
\end{align}
Our goal is to find the minimum value of $f(a,b,d)$. 

First, we show that $f(a,b,d)$ is minimum at $d=0$ by analyzing
\begin{align}
\frac{\partial f(a,b,d)}{\partial d}=-8 a^2 + 8 a^3 + 4 b + 8 a^2 b - 8 b^2 + 8 b^3 + 
 4 a (1 - 2 b + 2 b^2) + 2 (4 a^2 + 4 b^2) d.
\end{align}
One can check that $\frac{\partial f(a,b,d)}{\partial d}\ge0$ for $a,b>0$ and $d\ge0$. Thus, $f(a,b,d)$ is minimum at $d=0$. Note that $\frac{\partial f(a,b,d)}{\partial d}=0$ at $a=b=0$. In this case, $f(0,0,d)=1$.

Next, we consider 
\begin{align}
f(a,b,0)&=\left(2 a^2+2 a (b-1)+2 b^2-2 b+1\right)^2,\\
\frac{\partial f(a,b,0)}{\partial b}&=4 (-1 + a + 2 b) (1 + 2 a^2 + 2 a (-1 + b) - 2 b + 2 b^2).
\end{align}
$\frac{\partial f(a,b,0)}{\partial b}$ becomes zero at $b=(1-a)/2$, where $a+(1-a)/2=(1+a)/2\le1$ for $0<a\le1$. There is no real solution of $b$ for $1 + 2 a^2 + 2 a (-1 + b) - 2 b + 2 b^2=0$. We also obtain
\begin{align}
\frac{\partial^2 f(a,b,0)}{\partial b^2}|_{b=(1-a)/2}=4 - 8 a + 12 a^2>0.
\end{align}
Thus, $f(a,b,0)$ is minimum at $b=(1-a)/2$.

Finally, we consider 
\begin{align}
f(a,(1-a)/2,0)=1/4 (1 - 2 a + 3 a^2)^2.
\end{align}
Its minimum value is $1/9$ at $a=1/3$. Therefore, the minimum value of $Z_2^{(3)}$ is 1/9 at $a=1/3, b=(1-a)/2=1/3, c=1-a-b-d=1/3, d=0$. This state is W-state
\begin{align}
|\text{W}\rangle=\frac{1}{\sqrt{3}}|001\rangle+\frac{1}{\sqrt{3}}|010\rangle+\frac{1}{\sqrt{3}}|100\rangle,
\end{align}
and the maximum value of $S^{(3)}_{2}$ in W-class is $S^{(3)}_{2 \; {|\text{W}\rangle}}=\frac{1}{2}\log 9$.

\section{L-entropy and tripartite logarithmic negativity}\label{App:TripartiteNegativity}
A curve of L-entropy, which has been proposed by \cite{Basak:2024uwc}, also has a similar behavior to genuine multi-entropy as in Figure \ref{fig:GMEn2Curve}. L-entropy $l_{ABC}$ in tri-partite systems is defined by
\begin{align}
l_{ABC}&:=[l_{AB}l_{BC}l_{CA}]^{1/3},\label{LE}\\
l_{AB}&:=\text{Min}\{2S^{(2)}(A),2S^{(2)}(B)\}-S_{R}^{(1,1)}(A:B).
\end{align}
Assuming the holographic correspondence, a curve of L-entropy was computed by the bulk area in a three-boundary wormhole as shown in Figure 17 of \cite{Basak:2024uwc}, which is similar to our black hole multi-entropy curve of genuine multi-entropy. 

Note that $l_{AB}$ does not have the additive property (\ref{additive}), and thus $l_{ABC}$ of a triangle state $\ket{\Delta}_{ABC}= \ket{\psi_1}_{A_RB_L}\ket{\psi_2}_{B_RC_L}\ket{\psi_3}_{C_RA_L} $ can be nonzero. To see it, let us compute $l_{AB}$ of $\ket{\Delta}_{ABC}$. By using the additive property of $S^{(2)}$ and $S_{R}^{(1,1)}$, and 
\begin{align}
    S_{R}^{(1,1)}(A:B)_{|\psi_1\rangle_{A_RB_L}}&=2S^{(2)}(A)_{|\psi_1\rangle_{A_RB_L}}=2S^{(2)}(B)_{|\psi_1\rangle_{A_RB_L}},\\
    S_{R}^{(1,1)}(A:B)_{|\psi_2\rangle_{B_RC_L}}&=S_{R}^{(1,1)}(A:B)_{|\psi_3\rangle_{C_RA_L}}=0,
\end{align}
we obtain
\begin{align}
&\;\;\;\;\;l_{AB \; \ket{\Delta}}\notag\\
&=\text{Min}\{2S^{(2)}(A)_{|\psi_1\rangle_{A_RB_L}}+2S^{(2)}(A)_{|\psi_3\rangle_{C_RA_L}},2S^{(2)}(B)_{|\psi_1\rangle_{A_RB_L}}+2S^{(2)}(B)_{|\psi_2\rangle_{B_RC_L}}\}\notag\\
&\;\;\;\;\;-S_{R}^{(1,1)}(A:B)_{|\psi_1\rangle_{A_RB_L}}\notag\\
&=\text{Min}\{2S^{(2)}(A)_{|\psi_3\rangle_{C_RA_L}},2S^{(2)}(B)_{|\psi_2\rangle_{B_RC_L}}\},
\end{align}
which can be nonzero. Similarly,  \begin{align}
l_{BC \; \ket{\Delta}}=\text{Min}\{2S^{(2)}(B)_{|\psi_1\rangle_{A_RB_L}},2S^{(2)}(C)_{|\psi_3\rangle_{C_RA_L}}\},\\
l_{CA \; \ket{\Delta}}=\text{Min}\{2S^{(2)}(C)_{|\psi_2\rangle_{B_RC_L}},2S^{(2)}(A)_{|\psi_1\rangle_{A_RB_L}}\},
\end{align}
can be also nonzero. Therefore, the tri-partite L-entropy $l_{ABC}$ (\ref{LE}) of the triangle state $\ket{\Delta}_{ABC}= \ket{\psi_1}_{A_RB_L}\ket{\psi_2}_{B_RC_L}\ket{\psi_3}_{C_RA_L} $ can be nonzero.

\begin{figure}[t]
    \centering
    \includegraphics[width=10cm]{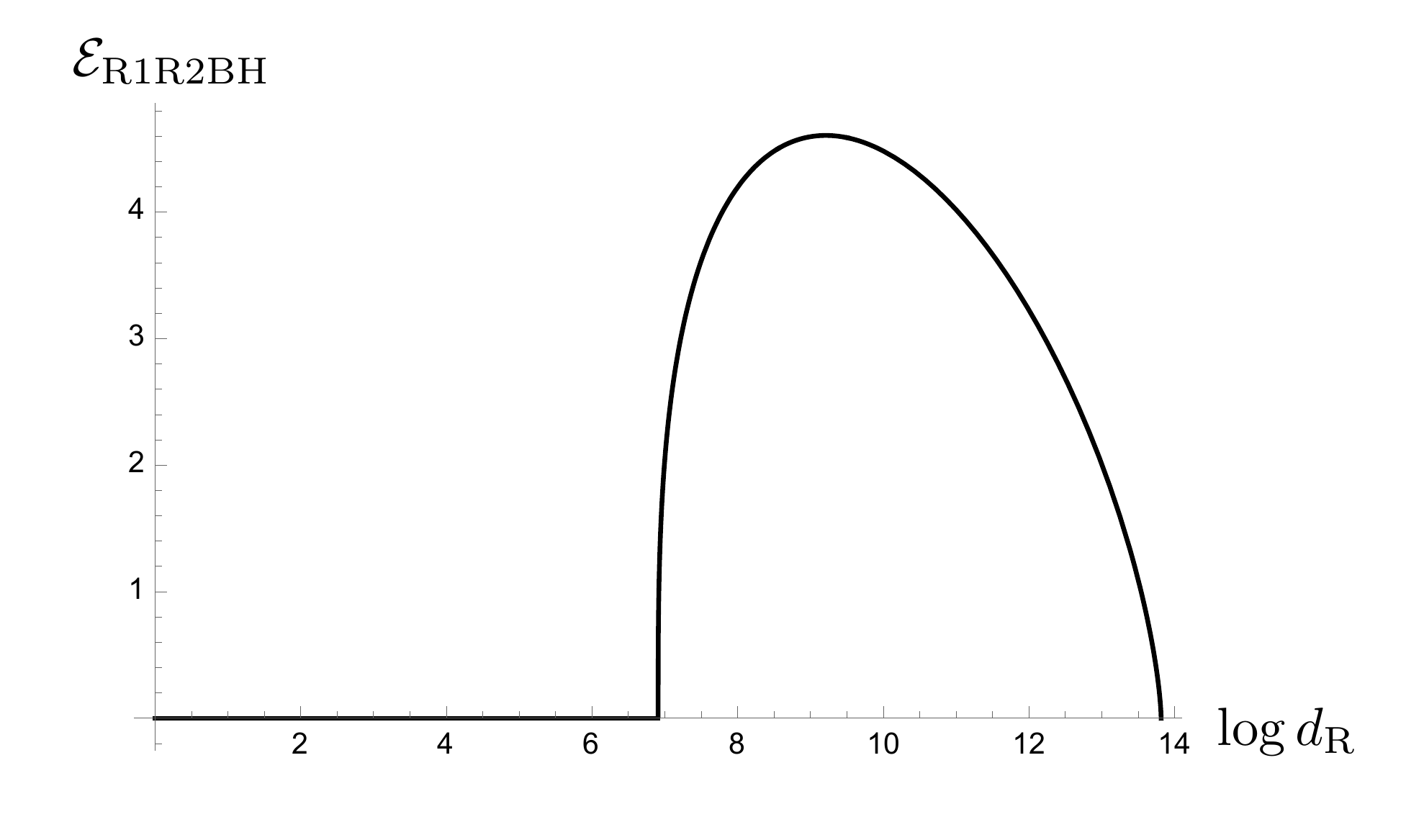}
    \caption{Asymptotic behavior of tripartite logarithmic negativity $\mathcal{E}_{\text{R1R2BH}}$. }
    \label{fig:TLN}
\end{figure}

By using the negativity \cite{Vidal:2002zz}, the tripartite negativity was defined by \cite{sabin2008}. In a similar way, we define tripartite logarithmic negativity $\mathcal{E}_{ABC}$ as follows.
\begin{align}
    \mathcal{E}_{ABC}:=[\mathcal{E}_{AB}\mathcal{E}_{BC}\mathcal{E}_{CA}]^{1/3},
\end{align}
where $\mathcal{E}_{AB}$ is the logarithmic negativity defined by
\begin{align}
\mathcal{E}_{AB}:=\log \|\rho_{AB}^{T_A}\|_1,
\end{align}
where $T_A$ means the partial transpose, and $\|\cdots\|_1$ is the trace norm. The asymptotic behavior of $\mathcal{E}_{AB}$ in a single random tensor model was derived by \cite{Bhosale:2012ldd, Lu:2020jza, Shapourian:2020mkc}. By using their results, we plot a curve of $\mathcal{E}_{\text{R1R2BH}}$ with the total dimension $d_{\rm Total} = d_{\rm BH}d_{\rm R}^{2}=10^{12}$ in Figure \ref{fig:TLN}, where the behavior is qualitatively similar to Figure 17 of \cite{Basak:2024uwc} for L-entropy $l_{R1R2B}$.

\section{The $\mathtt{q}$-partite information $I_{\mathtt{q}}$ curves for black hole evaporation}
\label{App:CforIq4}

\subsection{Setup for multipartite systems}

In this appendix, we study the $\mathtt{q}$-partite information $I_\mathtt{q}$, which is defined for $\mathtt{q}$ subsystems as follows:
\begin{align}
\label{defofIq}
I_\mathtt{q} (R_1, \cdots, R_{\mathtt{q}-1}, \text{BH}) := - \sum_{\sigma \subseteq \{R_1, \cdots, R_{\mathtt{q}-1}, \text{BH}\}} (-1)^{|\sigma|} S(\sigma),
\end{align}
where the sum is taken over all subsets $\sigma$ of $\{R_1, \cdots, R_{q-1}, \text{BH}\}$, and $|\sigma|$ denotes the number of elements in the subset $\sigma$. This particular linear combination of entanglement entropies is known to cancel all ultraviolet (UV) divergences \cite{Hayden:2011ag}. Note that despite being termed $\mathtt{q}$-partite information \cite{Mirabi:2016elb, Agon:2022efa}, it is constructed as a linear combination of bipartite entanglement entropies.
The total Hilbert space is given by
\begin{align}
\mathcal{H}_{\text{total}} = \mathcal{H}_{R_1} \otimes \mathcal{H}_{R_2} \otimes \cdots \otimes \mathcal{H}_{R_{q-1}} \otimes \mathcal{H}_\text{BH}.
\end{align}
This setup is motivated by scenarios in black hole evaporation where Hawking radiation is approximately isotropic. In such cases, it is natural to divide the radiation into $\mathtt{q}-1$ symmetric subsystems, as studied for instance in \cite{Iizuka:2024pzm}. Due to the symmetry, we assume the following:
\begin{align}
\dim(\mathcal{H}_{R_1}) = \cdots = \dim(\mathcal{H}_{R_{\mathtt{q}-1}}) = d_R, \quad \dim(\mathcal{H}_\text{BH}) = d_\text{BH},
\end{align}
so that the total dimension of the Hilbert space is
\begin{align}
\label{dtotal}
d_{\text{total}} = (d_R)^{\mathtt{q}-1} d_\text{BH}.
\end{align}
In our analysis, we keep $d_{\text{total}}$ fixed while varying $d_R$ and $d_\text{BH}$ accordingly.

We also assume that the entropy \( S_X \) of a subsystem \( X \) is determined only by its dimension \( d_X \), and follows the typical form for reduced subsystems in random pure states:
\begin{align}
\label{Pageformula}
S_X = \min \left[ \log d_X, \log d_{\text{total}} - \log d_X \right]
\end{align}

This form captures the behavior of the entanglement entropy in typical pure states and reflects the Page curve-like symmetry between a subsystem and its complement \cite{Page:1993df}.

Thus, we have the following properties:
\begin{itemize}
  \item The total system is pure, thus 
  \begin{align}
  \label{totalpureas}
  S_{R_1 \cdots R_{\mathtt{q}-1} \text{BH}} = 0 \,.
  \end{align}
  \item By symmetry between radiation subsystems, 
  \begin{align}
  \label{totalsymas}
  S_{R_1} = \cdots = S_{R_{\mathtt{q}-1}} \,, \quad S_{R_1 \text{BH}} =\cdots =S_{R_{\mathtt{q}-1} \text{BH}} \,, \quad  \mbox{etc}
  \end{align}
\end{itemize}

\subsection{$\mathtt{q}=3$ case: Tripartite information $I_3$}

First, let us consider the $\mathtt{q}=3$ case. 
The tripartite information is 
\begin{align}
I_3(R_1:R_2:\text{BH}) 
&= S_{R_1} + S_{R_2} + S_\text{BH} - S_{R_1 R_2} - S_{R_1 \text{BH}} - S_{R_2 \text{BH}} + S_{R_1 R_2 \text{BH}}
\end{align}

Using \eqref{totalpureas} and \eqref{totalsymas}, the expression simplifies to:
\begin{align}
\label{q3I3}
I_3 = 2 S_{R_1} + S_\text{BH}  - S_{R_1 R_2} - 2 S_{R_1 \text{BH}} \,.
\end{align}

We now evaluate each term using \eqref{dtotal}, {\it i.e.,} $d_{\text{total}} = d_R^2 d_\text{BH}$ and  \eqref{Pageformula}. 
\begin{enumerate}
\item Entropy of subsystem \( R_1 \):
\begin{align}
\label{SA1q3}
S_{R_1} = \min[\log d_{R}, \log d_{\text{total}} - \log d_{R}] = \min[\log d_{R}, \log d_{R} + \log d_\text{BH}] = \log d_{R}
\end{align}
since \( \log d_\text{BH} \ge 0 \).
\item Entropy of subsystem \( \text{BH} \):
\begin{align}
\label{SBq2}
S_\text{BH} = \min[\log d_\text{BH} , \log d_{\text{total}} - \log d_{\text{BH}}]= \min[\log d_\text{BH}, 2 \log d_{R}]
\end{align}
\item Entropy of subsystem \( R_1 R_2 \):
\begin{align}
S_{R_1 R_2 } =\min[2 \log d_R ,  \log d_{\text{total}} - 2 \log d_{R}]= \min[2 \log d_R, \log d_\text{BH}] = S_\text{BH}
\end{align}
due to \eqref{SBq2}, or equivalently, since the total system is pure.
\item Entropy of subsystem \( R_1 \text{BH} \):
\begin{align}
S_{R_1 \text{BH}} &= \min[\log d_{R} + \log d_\text{BH}, \log d_{\text{total}} - \log d_{R} - \log d_\text{BH}] \nonumber \\
&= \min[\log d_R + \log d_\text{BH}, \log d_R] = \log d_R 
\end{align}
since \( \log d_\text{BH} \ge 0 \). Then from \eqref{SA1q3}, we have \( S_{R_1 \text{BH}}  =  S_{R_1} \left(= S_{R_2} \right) \), which is consisntent since the total system  is pure. 
\end{enumerate}

Thus, substituting all into \eqref{q3I3}, we find:
\begin{align}
I_3 = 0
\end{align}
Therefore, the tripartite information $I_3$ always vanishes in this symmetric setup.

\subsection{$\mathtt{q}=4$ case: Quadripartite information $I_4$}

Similarly, let us consider $\mathtt{q}=4$ case. 
Using the symmetry \eqref{totalsymas}, the quadripartite information reduces to
\begin{align}
I_4(R_1:R_2:R_3:\text{BH}) &=  \sum_{i=1}^3 S_{R_i} + S_\text{BH} - \sum_{1 \leq i < j \leq 3} S_{R_i R_j} - \sum_{i=1}^3 S_{R_i \text{BH}}  \nonumber \\
& \qquad + \sum_{1 \leq i < j \leq 3} S_{R_i R_j \text{BH}} + S_{R_1 R_2 R_3}  - S_{R_1 R_2 R_3 \text{BH}}\\
&= 3 S_{R_1} + S_\text{BH} -3 S_{R_1 R_2} - 3 S_{R_1 \text{BH}}  + 3 S_{R_1 R_2 \text{BH}} + S_{R_1 R_2 R_3} 
\label{I4full}
\end{align}
Here we use \eqref{totalpureas} and \eqref{totalsymas}. 
Furthermore, the fact that the total system is pure combined with the symmetry  \eqref{totalpureas} implies 
\begin{align}
S_{R_1} = S_{R_3} = S_{R_1 R_2 \text{BH}} \, , \quad S_\text{BH} = S_{R_1 R_2 R_3}  \, ,\quad S_{R_1 \text{BH}} = S_{R_3 \text{BH}} =  S_{R_1 R_2}\, , \quad \mbox{etc}
\end{align}
These furthermore simplify \eqref{I4full} as 
\begin{align}
\label{I4full2}
I_4(R_1:R_2:R_3:\text{BH})  &= 6 S_{R_1} +2  S_\text{BH} -6 S_{R_1 \text{BH}}  \,.
\end{align}

We now evaluate each term using \eqref{dtotal}, {\it i.e.,} $d_{\text{total}} = d_R^3 d_\text{BH}$ and  \eqref{Pageformula}. 
\begin{enumerate}
\item Entropy of subsystem \( R_1 \):
\begin{align}
S_{R_1} = \min[\log d_{R}, \log d_{\text{total}} - \log d_{R}] = \min[\log d_{R}, 2 \log d_{R} + \log d_\text{BH}] = \log d_{R}
\end{align}
(since \( \log d_R \ge 0  \, , \,\log d_\text{BH} \ge 0 \))
\item Entropy of subsystem \( \text{BH} \):
\begin{align}
S_\text{BH}  &= \min[\log d_\text{BH}, \log d_{\text{total}}  - d_\text{BH} ] = \min[\log d_\text{BH}, 3 \log d_{R}] \nonumber \\
&=
\begin{cases}
3 \log d_R & \text{Before the Page time: when } d_\text{BH} \ge d_R^3 \\
 \log d_\text{BH} & \text{After the Page time:  when } d_\text{BH} \le d_R^3 \\
\end{cases}
\end{align}
The behavior changes at the Page time, defined by the condition \( d_\text{BH} = d_R^3 \).
\item Entropy of subsystem \( R_1 \text{BH} \):
\begin{align}
S_{R_1 \text{BH}} &= \min[\log d_{R} + \log d_\text{BH}, 2 \log d_{R}] \\
&= \begin{cases}
2 \log d_R  \qquad \qquad \, \text{Before the multi-entropy time: when } d_\text{BH} \ge d_R \\
\log d_R + \log d_\text{BH} \quad \text{After the multi-entropy time: when } d_\text{BH} \le d_R \\
\end{cases}
\end{align}
The behavior changes at the \textit{multi-entropy time}, as defined in \cite{Iizuka:2024pzm},\footnote{The multi-entropy time is independent of \( \mathtt{q} \).} which occurs when \( d_R = d_\text{BH} \), {\it i.e.}, when all subsystems have the same Hilbert space dimension.
\end{enumerate}

Thus, we find: \eqref{I4full2} reduces to 
\begin{align}
\boxed{
I_4 =
\begin{cases}
\qquad \qquad 0 & \text{Before the Page time: when } d_\text{BH} \ge d_R^3  > d_R\\
2 (\log d_\text{BH} - 3 \log d_R) & \text{After the Page time but before the multi-entropy time:} \\ \nonumber  &\qquad  \mbox{when} \quad d_R < d_\text{BH} \le d_R^3\\
\qquad -4 \log d_\text{BH} & \text{After the multi-entropy time (near the end of evaporation):} \\ \nonumber & \qquad \mbox{when} \quad  d_\text{BH} \le d_R < d_R^3\\
\end{cases}
}
\end{align}

\begin{figure}[t]
    \centering
    \includegraphics[width=8.5cm]{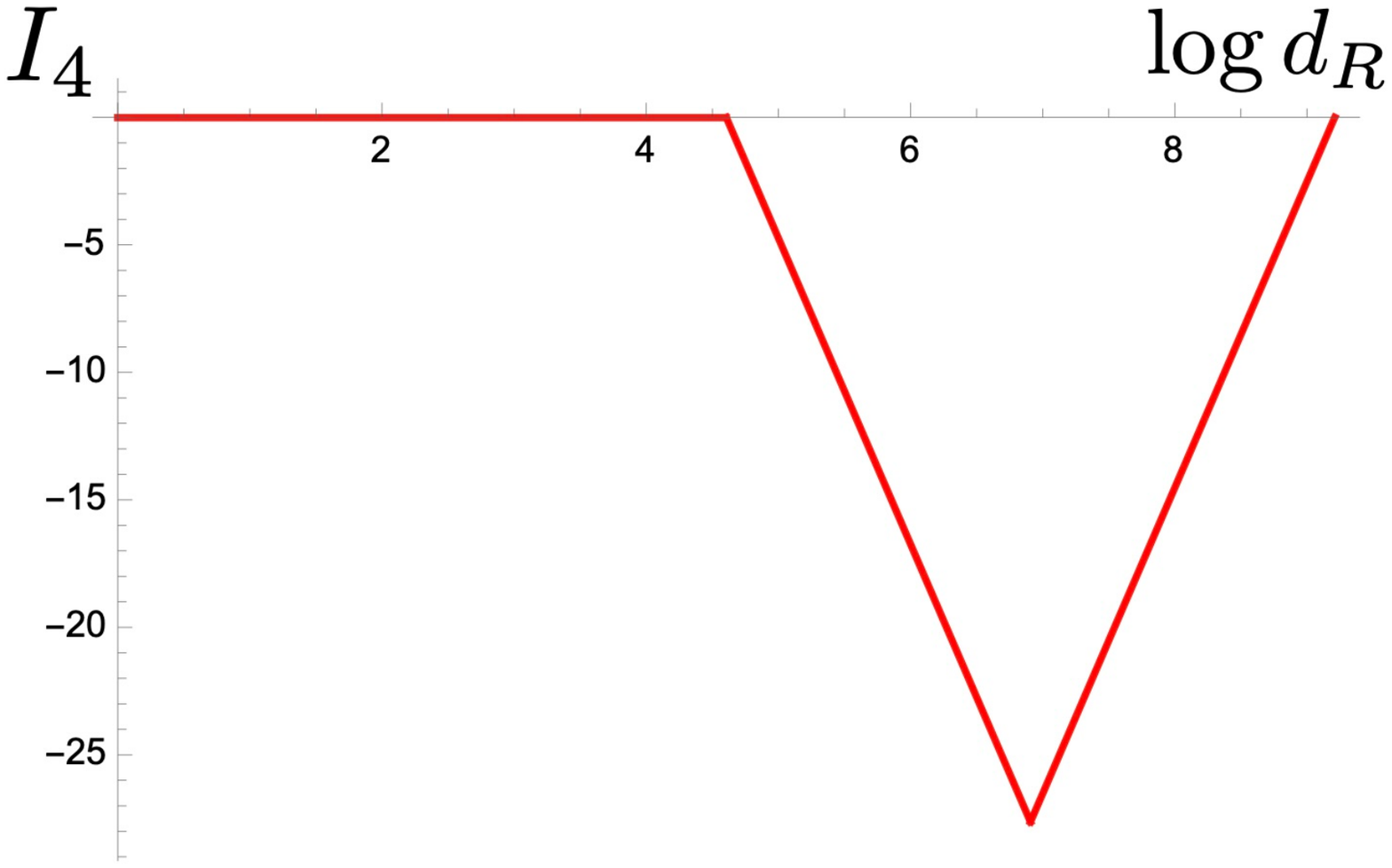}
    \caption{Asymptotic behaviors of $I_{\mathtt{q}=4}$ for a Haar-random state where $\dim\mathcal{H}_\text{R$_1$}=\dim\mathcal{H}_\text{R$_2$}=\dim\mathcal{H}_\text{R$_3$}=d_R$, $\dim\mathcal{H}_\text{BH}=d_\text{BH}$, and we fix $d_{\rm Total} = d_{\rm BH}d_{R}^{3}=10^{12}$. There are two time scales: (1) The Page time: $d_R^3=d_\text{BH}$, and (2) The multi-entropy time: $d_R=d_\text{BH}$.}
    \label{fig:App}
\end{figure}

\bibliography{Refs}
\bibliographystyle{JHEP}

\end{document}